\documentclass[12pt,preprint]{aastex}

 \newcommand{\msun}{$M_\odot$}
 \newcommand{\kms}{km s$^{-1}$}
 \newcommand{\numstars}{114}

\slugcomment{Accepted for publication in the Astrophysical Journal}

\shorttitle{Massive Binary Characteristics}
\shortauthors{Kobulnicky \& Fryer}
 
\begin{document}

\title{A New Look at the Binary Characteristics of Massive Stars}

\author{Henry A. Kobulnicky}
\affil{Department of Physics \& Astronomy, University of Wyoming,
Laramie, WY 82070} 
\email{chipk@uwyo.edu}

\author{Chris L. Fryer}
\affil{Theoretical Astrophysics, Los Alamos National Laboratory,
P.O. Box 1663, Los Alamos NM 87545}
\email{fryer@lanl.gov}

\begin{abstract}
We constrain the properties of massive binaries by comparing
radial velocity data on 114 early-type stars in the
Cygnus OB2 Association with the expectations of Monte
Carlo models.   Our comparisons test several
popular prescriptions for massive binary parameters
while highlighting the sensitivity of the best-fitting
solutions to the adopted boundary conditions.  We explore a
range of true binary fraction, $F$, a range of power-law
slopes, $\alpha$, describing the distribution of companion
masses between the limits $q_{low}$ and 1, and a range of
power-law slopes, $\beta$, describing the distribution of
orbital separations between the limits $r_{in}$ and
$r_{out}$.  We also consider distributions of secondary
masses described by a Miller-Scalo type initial mass
function (IMF) and by a two-component IMF that includes a
substantial ``twin'' population with $M_2 \simeq M_1$.
Several seemingly disparate prescriptions for massive binary
characteristics can be reconciled by adopting carefully
chosen values for $F$, $r_{in}$, and $r_{out}$.  We show
that binary fractions $F<0.7$ are less probable than
$F\geq0.8$ for reasonable choices of $r_{in}$ and $r_{out}$.
Thus, the true binary fraction is high.  For $F=1.0$ and a
distribution of orbital separations near the canonical
\"Opik's Law distribution (i.e., flat; $\beta=0$), the power
law slope of the mass ratio distribution is $\alpha=$-0.6 --
0.0.  For $F\simeq0.8$, $\alpha$ is somewhat larger, in the
range -0.4 -- 1.0.  In any case, the secondary star mass
function is inconsistent with a Miller-Scalo -like IMF
unless the lower end is truncated below $\sim$2--4 \msun.
In other words, massive stars preferentially have massive
companions.  The best fitting models are described by a
Salpeter or Miller-Scalo IMF for 60\% of secondary star
masses with the other $\sim40\%$ of secondaries having $M_2
\simeq M_1$, i.e., ``twins''.  These best-fitting model
parameters simultaneously predict the fraction of type Ib/c
supernovae to be 30--40\% of all core-collapse supernovae,
in agreement with recent observational estimates.
\end{abstract}


\section{Introduction}
\subsection{Significance and Theoretical Parameterizations of Binaries}

Massive binary stars are invoked as the progenitors of a
wide variety of astrophysical phenomena, from short and long
gamma-ray bursts \citep{fryer99}, type Ib/c and blue type II
supernovae \citep{pod}, to the fastest runaway O/B stars
\citep{blaauw} and the entire menagerie of binary systems
with compact remnants: X-ray binaries, millisecond pulsar
systems, and double neutron star systems (see
\citet{fryer98} for a review).  Whether binaries are
necessary for the formation of all such objects is a matter
of debate.  For instance, although many authors propose that
type Ib/c supernovae and short- and long-duration gamma-ray
bursts are produced in binaries, it is still argued that
their progenitors may be single stars \citep{hirschi}.
Observational studies can contribute to this debate by
measuring the frequency, separations, and mass ratios of
close massive binary systems.

Although debates on the role of binaries continue, it is
generally agreed that most of the progenitors of X-ray
binaries and double neutron star binaries in the Galactic
disk must arise from massive binary systems.  Population
synthesis models of these systems, however, adopt very
different properties for the progenitors' stellar
components.  These studies begin with four assumptions
regarding the binary characteristics: the distribution of
primary masses, the distribution of secondary masses, the
distribution of orbital separations, and the distribution of
eccentricities.  Population synthesis studies all choose
some analytic form for these distributions, generally basing
them upon observational constraints on the initial and final
binary system parameters.

One significant source of disagreement in these initial
distributions is the choice of the secondary star mass
function.  Typically, the primary star mass distribution is
chosen using a field initial mass function (IMF) 
\citep[e.g.,][]{scalo, salpeter}.  The secondary mass can be chosen through
the same IMF or, more commonly, from a distribution of the
secondary-to-primary mass ratios: $f(q)$, where $q \equiv
M_2/M_1$.  In this latter formalism, the distribution is
chosen to be proportional to the mass ratio to some power:
$f(q)\propto q^\alpha$ between the limits $q_{low}<q<1.0$.
Note that the power-law slope, $\alpha$, is not equivalent
to the power-law slope, $\gamma$, of the secondary star mass
distribution, commonly expressed $f(M_2)\propto M_2^\gamma$.
The relationship between $f(q)$ and $f(M_2)$ is discussed by
\citet{tout}, and it is, in general, non-trivial.
 
Population studies disagree on the choice of $\alpha$,
primarily because the choice of $\alpha$ strongly effects
the formation rate of key astrophysical objects.  High
values maximize the formation rate of double neutron star
binaries whereas low values maximize the formation rate of
low-mass X-ray binaries.  Generally, population synthesis
studies of low-mass X-ray binaries have assumed $\alpha$
values near -2.7 \citep{kalogera}, whereas studies of double
neutron star binaries tend to prefer a flat distribution
with $\alpha=0$
\citep{belczynski}.  Resolving this 
ambiguity is important because the choice of this value can
change formation rates of, for example, neutron star
binaries by nearly a factor of 100 \citep{fryer99}!  This is
the difference between advanced LIGO detecting 10 binary
inspirals per year versus 1 inspiral in 10 years.

The second critical parameter is the distribution of
orbital separations, $f(r)$.  This distribution is often
taken to be a power law between some inner and outer limits,
$r_{in}$ and $r_{out}$, such that $f(\log r)\propto (\log
r)^\beta$ where $\beta>0$ indicates a preference for large
orbital separations and $\beta<0$ indicates a preference
for small separations. 

\subsection{Observational Constraints on Mass Ratios}

Observational studies of binaries have reached widely
varying, even disparate, conclusions regarding these key
parameters (reviewed in \citet{abt83, larson}).
\citet{garmany} found $\alpha=0 - 1$ for a sample of 67
massive spectroscopic binaries, suggesting that the mass
ratio distribution is either flat or peaked toward massive
companions.  Among B2--B5 primaries,
\citet{abt90} report a secondary star mass distribution consistent
with the \citet{salpeter} IMF, corresponding to
$\alpha\simeq-2.3$.  In a spectroscopic study of solar type
field stars, \citep{duquennoy} conclude $\alpha\leq-2$ for
systems with $q>0.1$, consistent with secondaries drawn from
a field star IMF.  Adaptive optics imaging surveys for
companions of B and A stars in the Scorpius OB2 Association
\citep{kouwenhoventhesis, kouwenhoven05, shatsky}, however, find
the mass radio distribution to be well-characterized by
$\alpha$=-0.4 --- -0.5.  \citet{fisher} report a flat
($\alpha\sim0$) distribution of mass ratios among
single-lined spectroscopic binaries (SB1s) in the solar
neighborhood, while double-lined spectroscopic binaries
(SB2s) show a peak near $q\sim1$.  Arguments abound over
whether this $q=1$ peak is real or a selection effect
arising from the fact that SB2s are only identified as such
if the mass and luminosity ratios are near unity
\citep{lucy, halbwachs03}

Several investigations have concluded that there are two
populations of secondaries leading to a bimodal distribution
of mass ratios.  In summarizing data from the {\it Index
Catalog of Visual Binary Stars}
\citep{ids} \citet{hogeveen90, hogeveen92} and the 
6th {\it Catalog of the Orbital Elements of Spectroscopic Binary
Stars} \citep{dao8} found that $\alpha\leq-2 $ for systems
with mass ratios $q>q_0$ where $q_0\simeq0.3$.  This is
roughly equivalent to secondary masses being drawn from the
field star initial mass function
\citep{salpeter, scalo, kroupa91}.   For extreme mass ratio
systems, $q<0.3$, \citet{hogeveen92} finds a flat
distribution (i.e., $\alpha\simeq 0$).  Furthermore, the
turnover point, $q_0$, may vary inversely with primary mass,
but the \citet{hogeveen92} sample did not include O stars.
Using radial velocity data of solar-type F7-K stars with
periods $P<10$ yrs, \citet{halbwachs03} find a broad peak in
the mass ratio distribution near $q=0.5$ and a sharp peak
near $q=0.8$.  In a spectroscopic study of low-mass stars,
\citet{goldberg} report a peak in the mass ratio
distribution near $q=0.2$ and a smaller peak near $q=0.8$.
From a spectroscopic survey of eclipsing OB binaries in the
$SMC$ \citep{harries, hilditch} \citet{ps} conclude that
massive primaries have two populations of companions: a
``twin'' population with $M_2>0.95 M_1$ comprising 45\% of
binaries, and a population drawn from a flat mass
distribution ($f(q)=constant$ corresponding to $\alpha$=0)
comprising 55\%.  This report echos earlier findings among
low-mass stars that short-period ($P<40$ d) binaries have a
``twin'' component comprising 10\% -- 50\% of the total
population \citep{lucy, tokovinin}.  \citet{lucy06}
reassessed the \citet{hogeveen92}
study using data from the 9th {\it The Ninth Catalogue of 
Spectroscopic Binary Orbits}  
\citep{pourbaix} and concluded that the data support
an excess of  $q\simeq1$ ``twin'' systems.

\subsection{Observational Constraints on Orbital Separations}

\citet{duquennoy} use a radial velocity survey of G dwarf
primaries to conclude that the distribution of orbital periods
for nearby binaries is broad and approximately Gaussian in
$\log P$ having $\overline{\log P}=4.8$ and $\sigma_{\log
P}=2.3$ with $P$ in days.  Therefore, the distribution of
orbital separations should be approximately log-normal as
well, with a mean of $\sim25$ $AU$.  This result contrasts
with others that indicate the distribution of orbital
separations is flat in log space, $f(\log r)\propto (\log
r)^0$, corresponding to $f(r)\propto (r)^{-1}$ \citep[i.e.,
\"Opik's law;][]{opik}.
\citet{kouwenhoventhesis} and \citet{shatsky} measured
separations $r>30$ $AU$ for visual binaries in Sco~OB2 
and find a flat distribution consistent with \"Opik's
law.  \citet{lepine} report a flat distribution of
separations among {\it Hipparchos} common proper motion
stars in the solar neighborhood. 

\subsection{Observational Constraints on Binary Fraction}

Estimates for the binary fraction, $F$, among massive stars
vary widely, and such estimates are always lower limits
given that very close, very distant, and very low-mass
companions are difficult to detect.  Here, we use $F$ to
mean the fraction of all primary stars having exactly one
secondary.  We do not explicitly consider triple and higher
order systems, even though \citet{garcia} report that a
significant fraction of massive close binaries may have
multiplicity $>2$. \citet{abt83, abt90, poveda82} report
binary fractions of at least $F=0.7$, and suggest that the
binary fraction may be close to 100\% after corrections
for incompleteness.  Among O stars in clusters and
associations, \citet{gies} report binary fractions of
$F=$26\% -- 55\%.  \citet{mason} find binary fractions of at
least $F=$59\% -- 75\% for O stars in clusters and
associations. \citet{garcia} tabulate multiplicity
statistics for O stars in a variety of cluster environments,
and they conclude that the binary fraction varies between
$F=$14\% and 80\%, with denser clusters having lower binary
fractions.

\subsection{Scope of this Paper}

Despite a wealth of observational data on parameters of
binary systems, there is little apparent agreement, leaving
reasonable doubt as to whether the results are reconcilable.
Much of the disagreement can be explained by observational
and sample selection biases.  Both spectroscopic radial
velocity surveys and direct imaging surveys for companions
of massive primaries become incomplete at low mass ratios
below $q\sim0.25$.  Secondaries in close, extreme mass ratio
systems with massive primaries are presently undetectable
with either technique and are likely to remain so in the
near future.  Thus, estimates of $F$, $f(r)$, and $f(q)$ at low $q$
must be made by extrapolating some functional form of the
distribution based on data from systems with larger $q$
and/or large $r$.  Radial velocity surveys such as
\citet{garmany, abt90, duquennoy, halbwachs03, goldberg,
harries, fisher} preferentially detect close, short-period
companions ($a\leq few\times10$ $AU$) while imaging surveys
such as \citet{shatsky, kouwenhoven05, kouwenhoventhesis,
lepine} preferentially detect more distant, long-period
companions.  If the fundamental parameters of binaries are
different for close versus distant binaries
\citep{abt90,mazeh,mason}, then the variety of reported
results might be partially understood in terms of these
observational selection effects (see Fig~1 of
\citet{mason}; also \citet{tout}).    Furthermore,
many studies do not explicitly describe the limits,
$r_{in}$, $r_{out}$ over which the results are valid,
complicating comparisons between surveys.

The statistics on companions to massive stars are especially
uncertain as there have been few studies of massive
binaries since the survey of \citet{garmany}.  For massive
primary stars, the small number of broad spectral features,
coupled with rotational broadening, make radial velocity
measurements difficult, and the
\citet{garmany} sample may have missed binaries below
velocities of 50\,km\,s$^{-1}$.  More than half of their
sample consisted of double-lined spectroscopic binaries,
thereby introducing a bias in favor of nearly equal mass
components with large relative radial velocities.  
About half of the sample consisted of evolved O stars,
selected on a magnitude-limited basis from throughout the
sky.  Given the selection biases and lack of a common origin
and age for this sample, the conclusions of \citet{garmany}
may not apply to the general population of zero-age massive
binaries.  Similarly, the conclusions of \citet{ps}
are based upon a small number of eclipsing systems in the SMC, and 
are, therefore, most sensitive to systems of equal mass
and radius so that no firm conclusions can be drawn \citep{lucy06}.  

In this paper, we analyze data from the radial velocity
survey of 120 O and early-B stars in Cygnus OB2 \citep{kiminki}
to help resolve these disparate conclusions regarding the
binary characteristics of massive stars.  The new data more
than double the sample size of the \citet{garmany} study,
include massive stars down to early B spectral types, and
often involve more epochs of observations.  Although orbital
parameters for a large number of systems are not yet
available, we use the observed distribution of radial
velocities, in conjunction with Monte Carlo models, to test
several popular formulations of $F$, $f(q)$, and $f(r)$ from
the literature \citep{ps,hogeveen92,garmany}.  Using only
the raw velocities of a large sample of stars from one young
OB association mitigates many of the selection and
evolutionary effects responsible for conflicting results
from more general binary surveys and catalogs
\citep{tout}. 

We find that we are able to constrain the probable values
for $\alpha$ and $\beta$ to a narrow region of parameter
space, subject to reasonable assumptions about the boundary
conditions of the mass ratios and orbital separations.  Only
select combinations of $\alpha$, $\beta$, $r_{in}$,
$r_{out}$, and $F$ can reproduce the levels of velocity
variation seen in the data.  Some canonical
parameterizations for binary characteristics are
inconsistent with the data and can be ruled out.  Our
immediate intent is to provide improved input parameters for
population synthesis models that predict the rates of
energetic phenomena in the Milky Way and at cosmic
distances.  We also expect that these results will have
implications for the formation scenarios of massive stars
which are currently under debate \citep{bonnell98, wolfire, krumholz05}.

\section{The Data}

\citet{kiminki} report on a 6--year radial velocity survey
of 120 early-type probable members of the Cygnus OB2
Association.  Cygnus OB2 was chosen as a target for long
term radial velocity study to determine the binary fraction
and mass ratio distribution among massive binaries {\it in a
populous young OB association where the stars are of
comparable age and stellar evolutionary effects are
minimized.}  However, even in the $\sim$2--3 Myr old Cygnus
OB2, the presence of a variety of evolved stars suggests
that the star formation process was non-coeval \citep{mt,
hanson}. 

Spectroscopic observations from the Lick Observatory Shane
telescope, the Keck Observatory Keck I
telescope\footnote{Some of the data presented herein were
obtained at the W.M. Keck Observatory, which is operated as
a scientific partnership among the California Institute of
Technology, the University of California and the National
Aeronautics and Space Administration. The Observatory was
made possible by the generous financial support of the
W.M. Keck Foundation.}, the Kitt Peak National
Observatory\footnote{NOAO is the national center for
ground-based nighttime astronomy in the United States and is
operated by the Association of Universities for Research in
Astronomy (AURA), Inc., under cooperative agreement with the
National Science Foundation.} WIYN\footnote{The WIYN
Observatory is a joint facility of the University of
Wisconsin, Madison, Indiana University, Yale University, and
the National Optical Astronomy Observatories.  } telescope,
and Wyoming Infrared Observatory 2.3 m telescope (WIRO) over
the period 1999 July to 2005 October were used to measure
radial velocity variations of stars earlier than spectral
type B3. Sample stars were observed on at least 3 epochs
over this period.  Some stars have as many as 18 epochs of
data.  The mean and median number of epochs are 7.6 and 7.0
respectively.  The time sampling is irregular, as imposed by
telescope schedules and weather conditions.  Sampling is
additionally modulated by the observing season for Cygnus
(June through November).
The data were reduced and analyzed using cross-correlation techniques,
as described by \citet{kiminki}. 

\begin{figure}
\epsscale{1.0}
\plotone{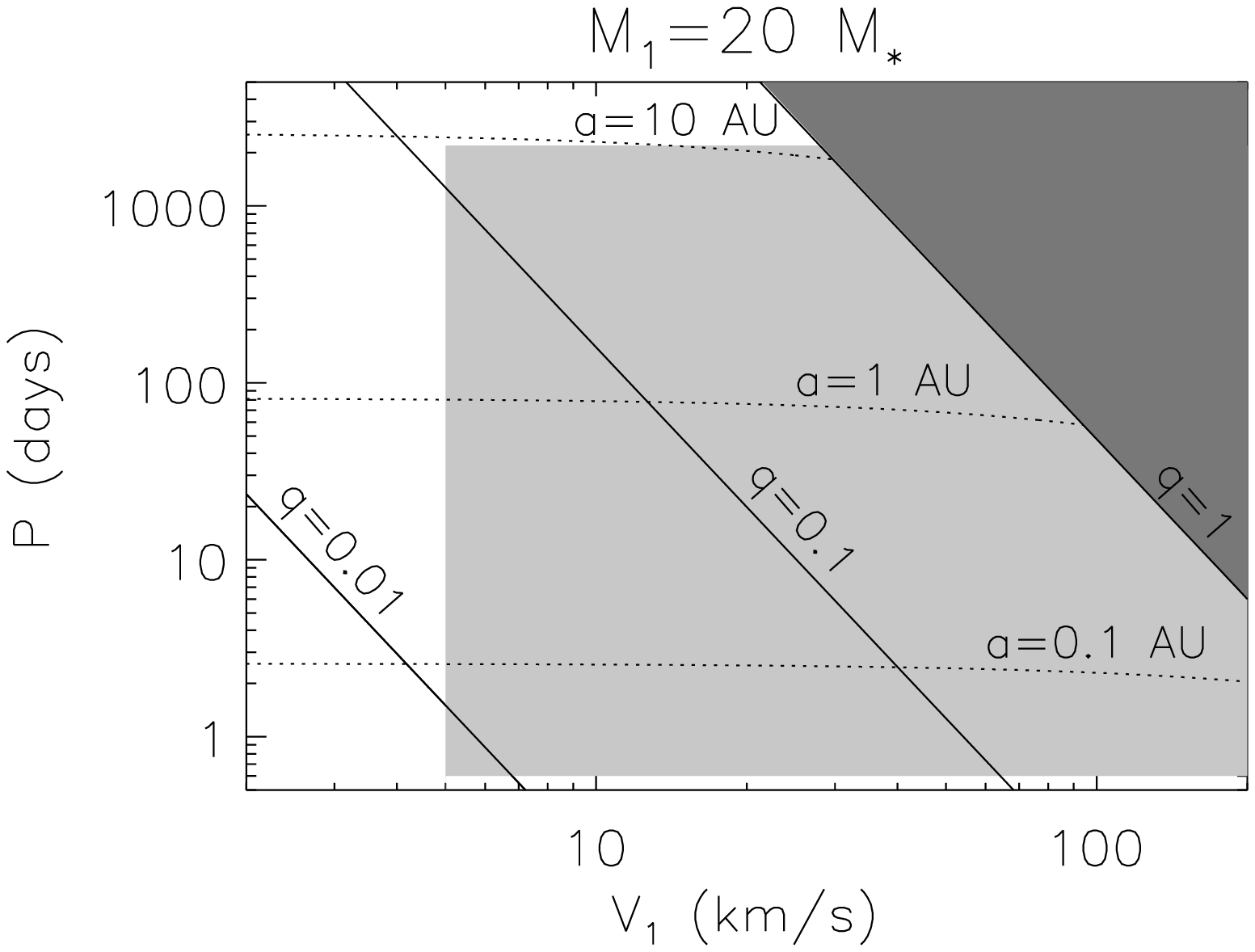}
\caption{Primary star velocity semi-amplitude versus orbital period 
for binary systems with a 10 \msun\ primary.  Solid lines indicate the
loci of systems with equivalent mass ratios, $q\equiv M_2/M_1$=1.0,
0.1, 0.01.  Dotted lines indicate loci of systems with common orbital
separations, $r=0.1, 1.0, 10$ $AU$.  The light gray region indicates the
portion of parameter space sampled in this survey.  Namely, the data
are sensitive to primary velocity amplitudes $\geq10$ \kms, periods
shorter than $\sim2 \times 6$ years, and orbital separations $\leq
2\times 10$ $AU$.  \label{params}}
\end{figure}

Figure~\ref{params} shows the analytical two-body
relationship between velocity semi-amplitude and orbital
period for binary systems with a 10 \msun\ primary.  Solid
lines indicate the loci of systems with equivalent mass
ratios, $q\equiv M_2/M_1$=1.0, 0.1, 0.01.  Dotted lines
indicate loci of common orbital separations, $r=0.1, 1.0,
10$ $AU$.  The dark gray area in the upper-right demarcates the
prohibited region of parameter space where $q\geq1$.  The
light gray area indicates the portion of parameter space
sampled in this survey.  The data are sensitive to primary
velocity semi-amplitudes $\geq 5-10$
\kms, periods shorter than $\sim2\times 6$ years and 
orbital separations $\leq\sim2\times10$ $AU$.

The data presented in \citet{kiminki} show at least several
dozen stars with strong radial velocity variations at the
level of $>20$ \kms\ and dozens of additional
velocity-variable candidates with semi-amplitudes down to
the survey sensitivity of $\sim5-10$ \kms.  There are also
$\sim6$ probable double-lined spectroscopic binaries, but
the individual components are poorly resolved in most of our
data, so these systems are, for present purposes, treated as
single-lined binaries.  For this analysis, we removed from
the main sample of 120 objects (Table~5 of \citet{kiminki})
six stars later than B4.  The remaining sample used for
analysis is \numstars\ objects.  These include the stars
identified as probable members by \citet{mt} plus additional
early B type members identified photometrically by
\citet{mt} and classified by \citet{kiminki}.  Given the
significant uncertainties in the photometric distances, we
have undoubtedly included a few foreground and background
stars.  The distribution of photometric distances (Figure~5
of \citet{kiminki}) is nearly Gaussian with only a couple of
outliers, suggesting that the sample does not suffer
significant contamination from field stars.  Including a few
field objects in the 114-star sample will not measurably
impact the conclusions, even if the binary characteristics
of field OB stars are different than the Cygnus OB2 members.

\begin{figure}
\epsscale{1.0}
\plotone{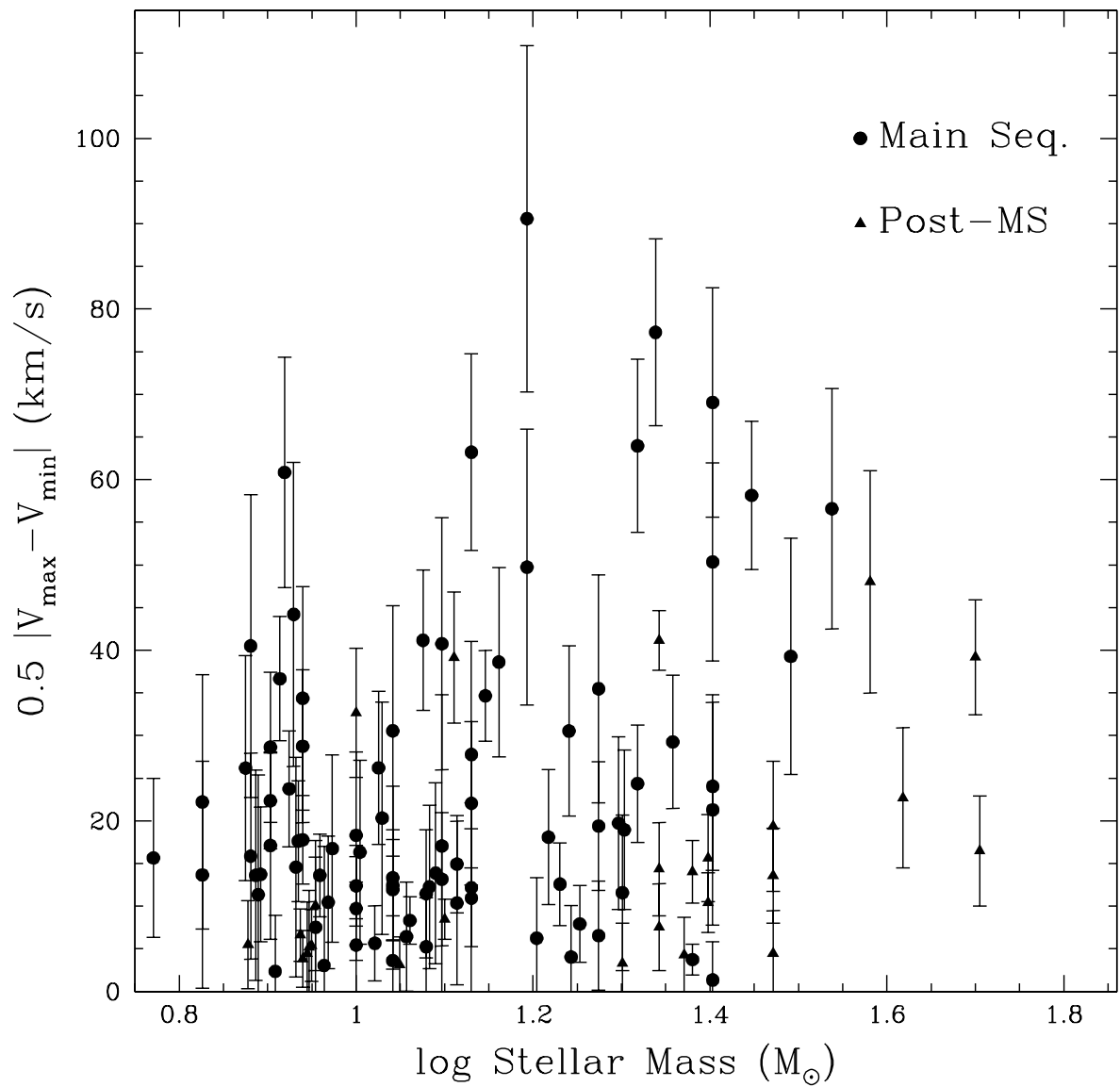}
\caption{Primary star spectroscopic masses 
versus observed velocity semi-amplitudes, $V_h\equiv 0.5|V_{max}-V_{min}|$,
for the \numstars\  sample stars. Typical uncertainties are 5--15 \kms.  
    \label{vels}}
\end{figure}

Figure~\ref{vels} plots the log of the assigned
spectroscopic masses \citep{martins, hm}
for each primary star versus the quantity $V_{h}\equiv 0.5
|V_{max}-V_{min}|$, where $V_{max}$ and $V_{min}$ are the
maximum and minimum observed velocity for a given star.
$V_{h}$ is a measure of the velocity semi-amplitude of the
primary, albeit an imperfect one because the velocity curves
are not generally well sampled at all phases. We 
regard $V_h$ as a lower limit on the true projected velocity
semi-amplitude.  Filled circles in Figure~\ref{vels} denote
main sequence OB stars while triangles denote evolved stars.
The error bars show the mean uncertainty in the velocity
measurements over all epochs for a given star.  A small
dispersion has been added to the masses in a few heavily
populated mass bins to improve clarity.  One O3If star
(MT457 in the nomenclature of \citet{mt}) at $M=80$ \msun\
falls outside the maximum plot range.  The typical velocity
precision is $\sim5-15$ \kms, and it varies somewhat with
source luminosity and observatory/instrument combination
used.  For example, the radial velocity uncertainties
obtained with the WIYN telescope + Hydra spectrograph are
often better than those obtained with the Lick 3 m telescope
+ Hamilton echelle spectrograph because of the higher signal
to noise ratio achievable with the former.
\citet{kiminki} provide a more complete description of these data 
and place a lower limit on the binary fraction at 30\% -- 42\%.  

\begin{figure}
\epsscale{1.0}
\plotone{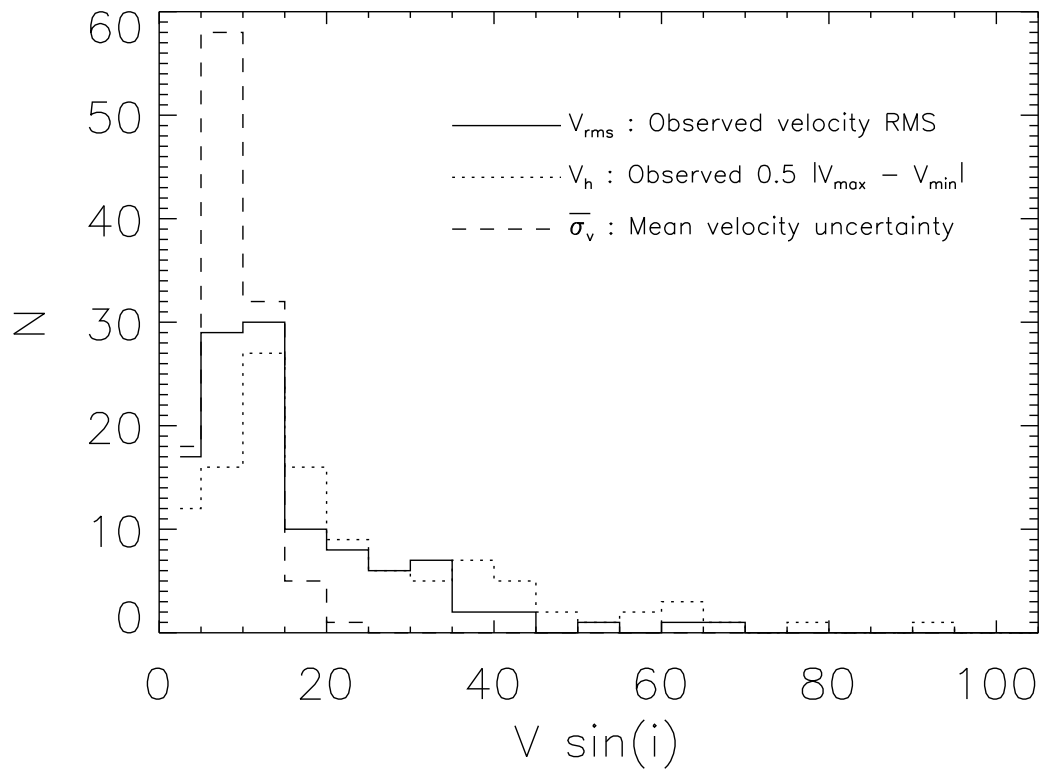}
\caption{The distribution of 
observed velocity dispersions, $V_{rms}$ (solid line),
velocity semi-amplitudes,  $V_h\equiv 0.5|V_{max}-V_{min}|$  (dotted line),
and the mean velocity uncertainties (dashed line) for the sample.
    \label{obsdist}}
\end{figure}

Figure~\ref{obsdist} shows a histogram of the observed
velocities and uncertainties.  This Figure illustrates the
distribution of velocity dispersions, $V_{rms}$, calculated
from the multiple measurements of the \numstars\ OB stars
(solid line) along with the distribution of mean velocity
uncertainties, $\overline{\sigma}_v$ (dashed line).  The
dotted line shows the distribution of $V_h$.  The lowest
velocity bin from 0 to 5 \kms\ is sparsely populated because
observational errors scatter the data into higher velocity
bins.  The maximum observed semi-amplitudes fall mostly
between 10 and 40 \kms, with a significant tail toward
higher velocities out to $\sim$90 \kms.  The uncertainties
lie in the characteristic range 5 -- 15 \kms.

\begin{figure}
\epsscale{1.0}
\plotone{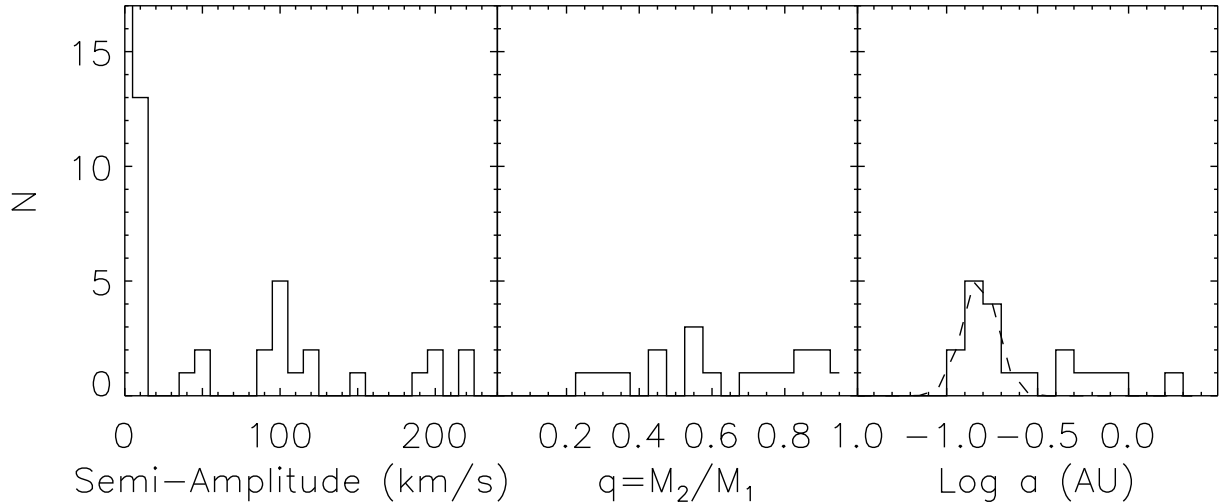}
\caption{{\it Left panel:} Histogram of the velocity semi-amplitudes 
for the O stars presented in Table 3 and Figure~1 of \citet{garmany}.
Note the differences compared to our Figure~\ref{obsdist}, namely a relative
lack of stars in the velocity range 20-80 \kms\ in the \citet{garmany} sample.
{\it Middle panel:} Histogram of mass ratios, $q$, for the 
O type binaries from \citet{garmany}.  The histogram is
consistent with a flat or rising distribution with increasing mass ratio.
{\it Right panel:} Derived semi-major axis distribution for the O type binary 
systems from \citet{garmany} assuming an average orbital
inclination of 60\degr.  The dashed line is a best fit Gaussian curve 
which peaks near $\log(r)=-0.8$ or $\sim0.15$ $AU$ with a tail toward larger separations.
    \label{garmany}}
\end{figure}

For comparison, Figure~\ref{garmany} shows the distribution
of properties for O-type systems from \citet{garmany}.  The
left panel shows a histogram of the velocity semi-amplitudes
for the new observations of O stars presented in Table~3 and
Figure~1 of \citet{garmany}.  Note the differences compared to our
Figure~\ref{obsdist}.  Most striking in the Garmany et
al. sample is the relative lack of stars in the velocity
range 20--80 \kms\ and the relative excess of stars in the
range 80--120 \kms.  Significant selection affects may drive
these differences.  The \citet{garmany} sample was magnitude
limited and contained stars at a variety of evolutionary
stages, including a large fraction of supergiants.  For such
systems, mass exchange and transfer of angular momentum
between orbital and rotational components may have altered
the original binary mass ratios and separations.  Therefore,
the \citet{garmany} sample may not be representative of the
general population of massive binaries at zero age.  The
middle panel of Figure~\ref{garmany} is a histogram of mass
ratios, $q$, for the O type binaries from Table~3 of
\citet{garmany}.\footnote{Here, we have included only the stars with
new observations  denoted by an asterisk in their Table~3.}
The histogram shows a preference for mass ratios near unity
and is consistent with a flat ($\alpha=0$) or rising
distribution ($\alpha>0$) with increasing mass ratio.  The
right panel of Figure~\ref{garmany} shows a derived
semi-major axis distribution for the O type binary systems
from Table~3 of \citet{garmany} assuming an average orbital
inclination of 60\degr.  The dashed line is a best fit
Gaussian curve which peaks near $\log(r)=-0.8$ or $\sim0.15$
$AU$.  This panel shows that the O-type binaries from Garmany
et al. are preferentially those with small orbital
separations, as might be expected since these are the
easiest to detect.  \citet{garmany} did consider whether the
lack of systems with small velocity amplitudes implied a
true absence of such systems or a merely limitation on their
detection.  They concluded that such low-amplitude systems
should have been detected as part of their survey and that
the preference for equal mass components was a real effect.

\section{Modeling Binary Characteristics}

\subsection{Monte Carlo Modeling of Radial Velocities}

We wrote a Monte Carlo code to simulate the radial velocity
variations expected for Cyg~OB2 primaries based upon
analytic prescriptions of binary mass ratios and orbital
separations.  We initially assumed a binary fraction of
$F=1.0$ so that every primary star has one companion.
Statistically, this means that single stars are treated as
systems with very low mass companions, very large
separations, or very low inclinations.  However, this is a
useful approach to obtain an initial lower limit on
$\alpha$.  For each of the 114 primary stars \citep{kiminki}
of mass $M_1$ and observed velocity variations characterized
by $V_h$ and $V_{rms}$, the Monte-Carlo code randomly assign
a secondary star of mass $M_2$ based upon a distribution of
mass ratios, $f(q)$ where $q\equiv M_2 / M_1$.  The mass
ratios are drawn from a population described by $f(q)
\propto q^\alpha$ where $\alpha$ is the power-law index of
the mass ratio distribution over the range $q_{low}<q<1$.
Initially, we adopt $q_{low}=0.02$ so that the lowest mass
secondaries for our least-massive $\sim$8 $M_\odot$
primaries are 0.16 $M_\odot$ and $\sim$1 \msun\ for the
handful of very massive 50 \msun\ stars. Such masses are
sufficiently low that the secondaries produce negligible
radial reflex motion in their primaries---well below the
velocity sensitivity of the data.  We consider values of
$\alpha$ between -3.0 and 2.0.  $\alpha>0$ describes a
secondary star mass distribution peaked toward the primary
mass.  $\alpha=0$ corresponds to a flat distribution of
secondary masses. The Monte-Carlo code assigns an orbital
separation, $r$, drawn from a distribution described by
$f[\log (r)] \propto [\log(r/r_{in})]^\beta$ with
$r_{in}<r<r_{out}$ and $r$ in A.U. \"Opik's Law ($\beta=0$) corresponds to
a parent population of orbital separations uniformly
distributed in $\log(r)$ between $log(r_{in})$ and
$log(r_{out})$.  $\beta=\pm3$ corresponds to distributions
peaked toward large/small separations, respectively.

Because the power-law slope, $\beta$, that best describes
the orbital separations of the data will depend upon the
chosen boundaries, $r_{in}$ and $r_{out}$, we explore a small
3x3 grid of values: $r_{in}=$0.020, 0.063, \& 0.200 $AU$ and
$r_{out}=$100, 1000, \& 10000 $AU$.  These inner limits
correspond to 4.2, 13, and 43 $R_\odot$.  For a
main-sequence B0 star, representative of our targets, these
inner limits correspond to 0.56, 1.7, and 5.8 $R_*$.
Thereby, the smallest two values for $r_{in}$ are firm lower
limits, representing separations where significant mass
transfer and common envelope evolution would take place.
Although these values are physically implausible in general,
we include them to illustrate the sensitivity of the
simulations to this parameter.  The third value, 0.20 $AU$,
is the largest reasonable choice for $r_{in}$, as it is
larger than several of the orbital separations for
short-period Cyg OB2 binaries reported in \citet{kiminki}
and \citet{garmany} ($\sim0.15~AU$).  The minimum adopted outer limit,
$r_{out}=100$ $AU$ is small compared to observed orbital
separations in wide binaries which may extend to several
thousand $AU$ \citep[e.g.,][]{kouwenhoven05, duquennoy}.  The
maximum adopted outer limit, $r_{out}=10000$ $AU$, is
sufficiently large that few systems are likely to remain
bound at such distances, especially in the dense stellar
environments where OB stars are born.
\citet{lepine} report that among nearby $Hipparchos$
binaries, the \"Opik's Law distribution of 
separations fits the data out to a maximum of $r=4000$ $AU$, beyond
which the number of companions drops, consistent with
gravitational disruption.

In the interest of minimizing free parameters, we assume
that the orbits have zero eccentricity.  The resulting
period and velocity semi-amplitude of the
simulated binary system is computed $N_{iter}$ times, where
$N_{iter}$ is typically 100 Monte-Carlo iterations.  For
each simulated system, the velocity curve is sampled at a
random phase angle $N_{obs}$ times, where $3<N_{obs}<18$,
the actual number of observations of a given star.  The code
assigns a random inclination angle, $i$, for each system,
where $i$ is generated by allowing the angular momentum
vector of the simulated system to lie anywhere on a sphere.
The mean observational velocity uncertainty for each
individual star, normally distributed about zero, is then
added to the simulated velocity.  As a test, we also
performed simulations using the actual uncertainties for
each measurement of each star and we found that the results
differed insignificantly from using the average
uncertainty.  

For each Monte Carlo run with a
given pair of $\alpha$ and $\beta$, the {\it simulated}
distribution of $V_h$ and $V_{rms}$ for \numstars\ stars is
compared to the {\it observed} distribution using a
two-sided K-S test.  The probabilities that the {\it
simulated} and {\it observed} distributions of $V_h$ and
$V_{rms}$ ($P(V_h)$ and $P(V_{rms})$, respectively)
are drawn from the same population are averaged over 100
Monte Carlo iterations and tabulated for later analysis.  We
found that both velocity metrics, $V_h$ and $V_{rms}$, yield
similar results and provide similar probabilistic
constraints on the binary characteristics. While $V_h$ is a
better measure of the true semi-amplitude, it is also less
robust than $V_{rms}$ against outliers or measurements with
large uncertainties.  Therefore, in subsequent discussion
and figures we use the average probability, $P_a$, obtained from the
arithmetic mean of $P(V_h)$ and $P(V_{rms})$.

In Section~4 we will discuss the contraints imposed by 
this modeling appraoch.  The Figures and discussion will show
that there is degeneracy between $\alpha$ and $\beta$---in general,
large $\alpha$ coupled with large $\beta$ reproduce the velocity data
approximately as well as small $\alpha$ coupled with small $\beta$.  
Additional constraints are needed.  One such constraint we consider
is that \"Opik's Law ($\beta=0$) obtains.  Another 
is the fraction of type Ib/c supernovae predicted by each model.
 
\subsection{Monte Carlo Modeling of Type Ib/c Supernva Rates}

The massive binary population can also be constrained by
observations of energetic phenomenae that arise from massive
binary progenitors.  A well-studied example is the fraction
of core-collapse supernovae that occur in stars that have
lost their hydrogen envelopes---type Ib/c supernovae.
\citet{pod} argued that most Ib/c supernovae are produced in
binary star systems where mass transfer removes most/all of
the hydrogen envelope, unveiling a type Ib/c supernova
progenitor.  \citet{pod} assumed that 37\% of binaries were
close binaries that would undergo some mass transfer.

The fraction of type Ib/c supernovae predicted by population
sysnthesis models must agree with observations {\it and}
must self-consistently produce the levels of velocity
variation seen in massive primaries, such as those in Cyg
OB2.  The fraction of type Ib/c supernovae relative to 
all core-collapse supernovae is 
$f_{Ib/c}\equiv N_{Ib/c}/ (N_{Ib/c}+N_{II})$.
\citet{cappellaro99} and
\citet{mannucci} give the observed fraction of
type Ib/c supernovae in Sbc/d spiral galaxies such as the
Milky Way as 15\%.  The more recent works of \citet{li} and
\citet{leaman} revise this estimate upward to 30\%$\pm$11\%.
Using binary population synthesis calculations, we can apply
these fractions to further constrain the parameter space in
our study. Any part of parameter space that produces a
fraction of type Ib/c supernovae inconsistent with the
observed fraction can be ruled out.

\begin{figure}
\epsscale{1.0}
\plotone{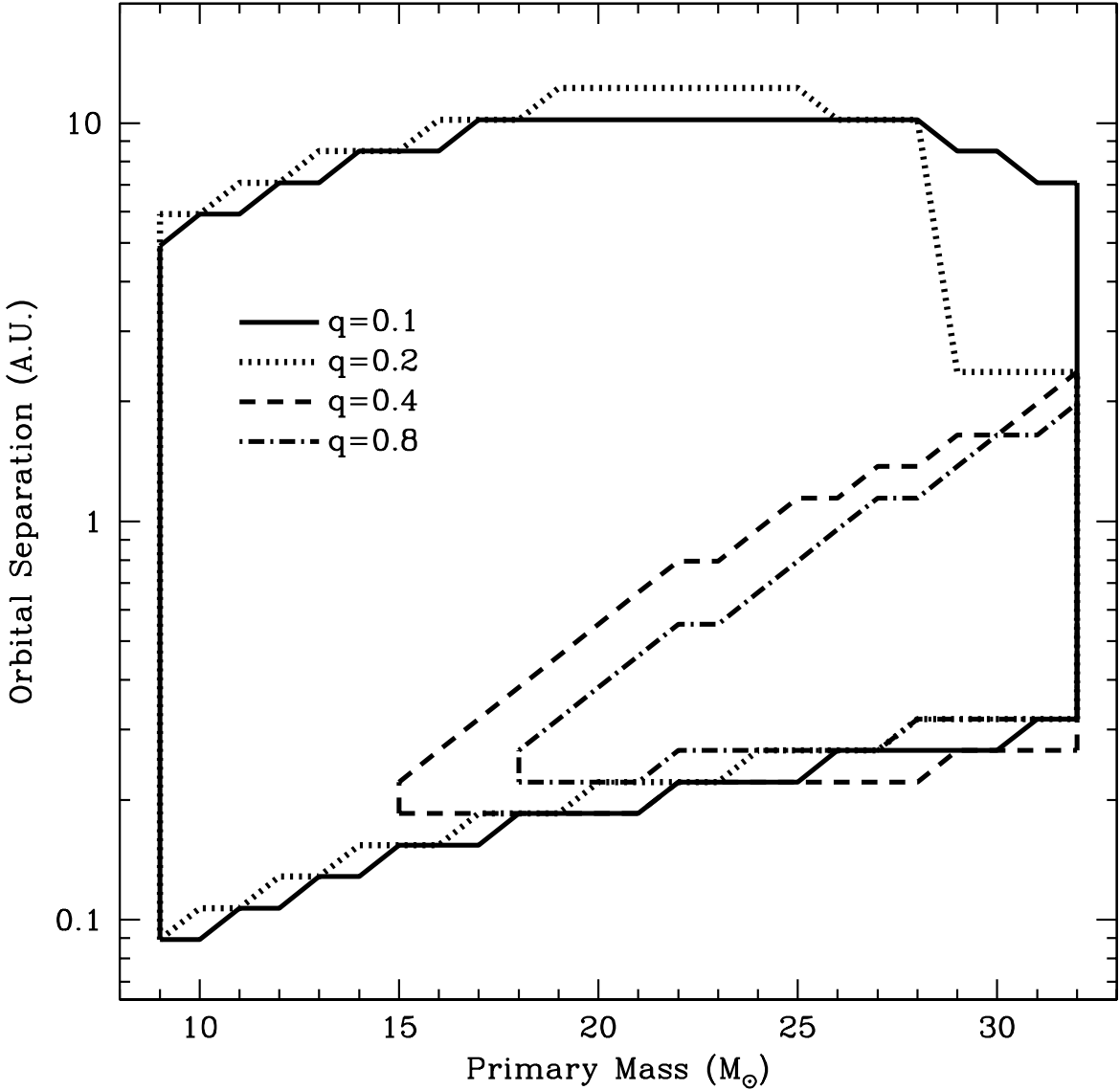}
\caption{Enclosed regions which denote the parameter space
of orbital separation and primary mass which produce type
Ib/c supernovae for 4 different values of the mass ratio, $q$.
Type Ib/c supernovae are preferentially made in systems that
undergo a common envelope phase, so the parameter space is
larger for more extreme mass ratios.  Above 32\,M$_odot$,
our wind prescription makes all stars (whether in
interacting binaries or not) lose enough mass to become type
Ib/c supernovae.
    \label{primary}}
\end{figure}

We used the binary synthesis code developed by
\citet{fryer98, fryer99} to calculate the total
core-collapse supernova rate and the number of type Ib/c
(stars that have lost their hydrogen envelopes at collapse)
supernovae.  Like Podsiadlowski (1992), we find that the
primary formation path for type Ib/c supernovae begins with
a close binary where the more massive primary envelops its
companion in a common envelope phase.  The companion spirals
into the envelope of the massive primary, ejecting the
hydrogen envelope.  After the common envelope phase, the
primary will continue to lose mass in a wind.  For this
paper, we will assume any star that loses its entire
hydrogen envelope has the potential to become a type Ib/c
supernova.  Because mass is lost in winds as well, even
systems (especially those at high metallicity) that merge in
the common envelope phase without ejecting the hydrogen
envelope can make type Ib/c supernovae.  Systems with more
extreme mass ratios (i.e., $q\leq0.2$) are more likely to go
through a common envelope phase and systems with such
low-mass companions dominate the formation rate of type Ib/c
supernovae.  Figure~\ref{primary} shows the range of primary
masses and orbital separations that form Ib/c supernovae for
4 separate values of the system mass ratio.  As $q$
increases, the system tends to undergo more stable mass
transfer without a common envelope phase.  Most of these
systems do not become type Ib/c supernovae.  These
simulations are based on our standard wind assumptions 
in which wind mass loss rates are set to 10\% of the \citet{woos} values.
Beyond 32\,M$_\odot$, these primary star winds will also
produce type Ib/c progenitors, regardless of companion mass.
However, this corresponds to only a small fraction of our
type Ib/c supernova rate.

The other requirement for a star to produce a type Ib/c
supernova is that it actually explode as a supernova.  We
assume any primary with an initial mass above $9\,M_\odot$
will produce a supernova.  Although the total rate of type
Ib/c supernova does depend on this choice, the dependence on
the type II rate is nearly identical and varying this value
in the $8-10\,M_\odot$ range leads to variations in the rate
of 2-3\% for the bulk of our parameter study.  We also
assume that stars that have helium cores in excess of
$11\,M_\odot$ collapse directly to black holes and don't
form supernovae.  Varying this value of the helium core mass
over the $8-15\,M_\odot$ range yields variations in the rate
of less than 1\%.  This is not surprising because the number
of such high-mass systems is quite small.

In this paper, we focus our binary population synthesis
studies on the mass distributions of the stellar components
and the orbital separation.  These distributions are varied
according to range of parameters studied in the analysis of
our observations and constitute over 10,000 different
population synthesis calculations each including 100,000
binaries for each calculation.  In all of these
calculations, we assume the eccentricity is zero.  If we
instead assume the eccentricity follows a flat distribution
between 0 and 0.8, we can increase the rate by 5\%.

We have already discussed the dependence of our ratios on
some of the many free parameters in population synthesis:
e.g. critical mass for supernova formation.  Let's briefly
discuss some of the remaining free parameters.  We assume a
single Maxwellian kick distribution with a mean value of
300\,km\,s$^{-1}$.  Because nearly all of the type Ib/c
supernovae are formed in hydrogen envelope stripping in the
common envelope inspiral of the companion star into the
massive primary, the rate does not depend strongly on the
choice of this kick.  We have varied the value of this mean
value from 100\,km\,s$^{-1}$ to 600\,km\,s$^{-1}$ as well as
using the bimodal kick distribution of Fryer et al. (1998)
and the rate does not change by more than 5\%.  For mass
transfer, we use the same formalism as Fryer et al. (1999)
with mass transfer parameters $\alpha_{\rm MT} =1.0$ and
$\beta_{\rm MT} =0.5$ and a common envelope efficiency
$\alpha_{\rm CE} =0.5$.  Varying these values for a wide
range of values (e.g. $0.2<\alpha_{\rm CE} <1.0$) yields
variations in the ratio of 10-30\%.

The final two uncertainties studied in our calculations are
our assumed stellar radii and stellar mass loss rates.  Here
again, we use the standard Fryer et al. (1998) values.  But
if we allow a factor of 2 variation (both increase and
decrease) in these values, the ratio of type Ib/c supernovae
to type II supernovae can vary by nearly 50
uncertainty is one of the dominant uncertainties in the
population synthesis of binary systems: the error
in the stellar radius.  If the stellar radii are smaller, we
produce fewer common envelope systems with a larger fraction
of the remaining common envelope systems leading to mergers
prior to the ejection of the hydrogen envelope.  This error
dominates our 50\% error.  Clearly we can vary the population
synthesis parameters to make many of the models fit to the
data.  But if we assume we know the stellar radii and take
standard values for common envelope efficiencies, our
calculation errors can be limited to 5--10\% effects,
allowing us to place strong constraints on the assumed
distributions of initial masses and separations.

\section{Results of Modeling}

\subsection{Binary Characteristics
for single power-law distributions of $q$ and $r$}

\begin{figure}
\plotone{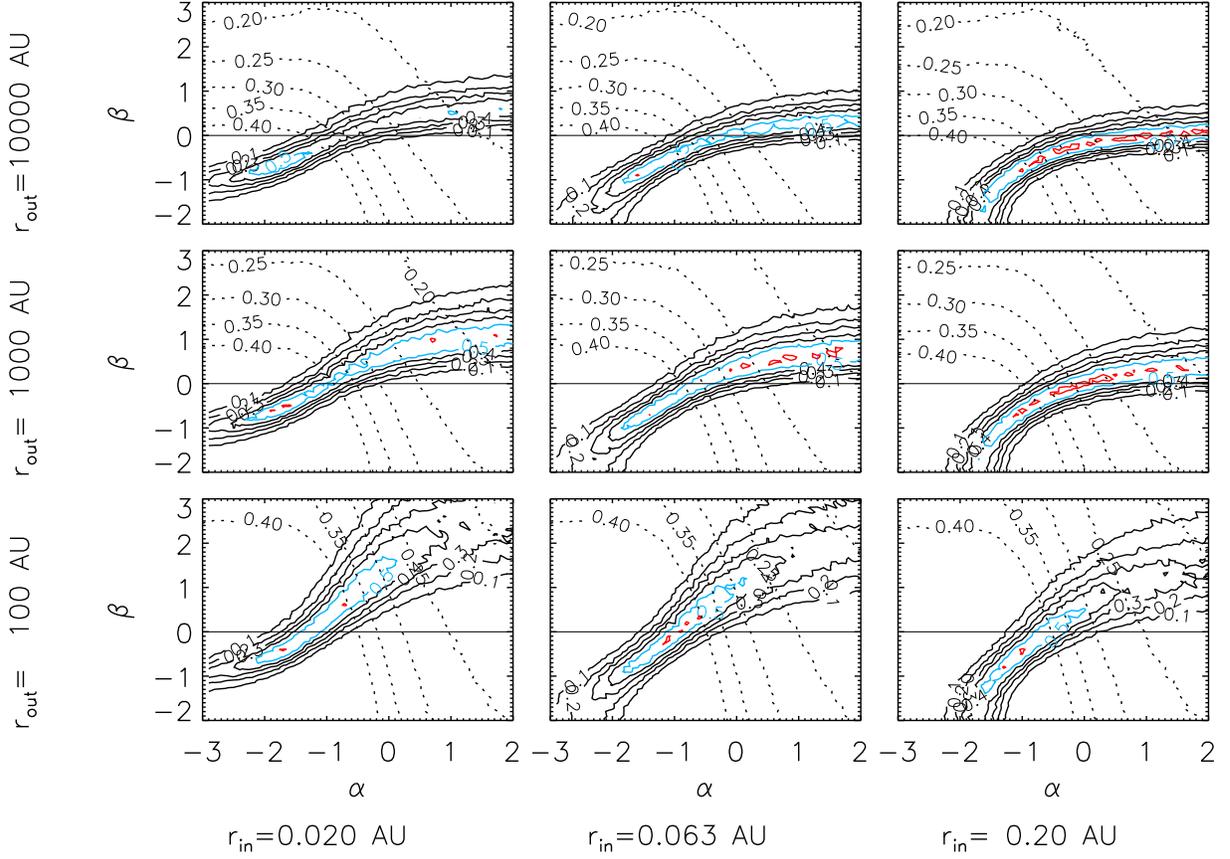}

\caption{Probability contours describing the agreement between the
the Monte Carlo simulations and the radial velocity data
for Cyg~OB2 (solid lines) as a function
of $\alpha$ and $\beta$ in each panel, for a range 
of minimum orbital radii, $r_{in}$ (columns),
and maximum orbital radii, $r_{out}$ (rows) with a
minimum mass ratio of $q=0.02$ and a binary fraction of 
$F=1.0$:  
$r_{in}=0.020~AU$ (left column), 
$r_{in}=0.063~AU$ (middle column), 
$r_{in}=0.200~AU$ (right column), 
and 
$r_{out}=10000~AU$ (upper row), 
$r_{out}=1000~AU$ (middle row), 
$r_{out}=100~AU$ (lower row).
The horizontal line in each panel marks the nominal $\beta=0$ (\"Opik's Law)
distribution of separations.
Contour levels probabilities $P_a=$ 10, 20, 30, 40, 50, and 60\% likelihood that the
simulations and the data are drawn from the same parent population.
The 50\% and 60\% contours are colored blue and red, respectively, in the
electronic edition of {\it The Astrophysical Journal.}
The most probable values exceed 60\% likelihood 
and lie along a narrow crescent-shaped ridge line.  
The dotted contours
depict the type Ib/c supernova fraction, $f_{Ib/c}=$0.10--0.40 
in increments of 0.05,
predicted by Monte Carlo binary population synthesis models \citep{fryer98, fryer99}.  
  \label{9panela}}
\end{figure}

Panels within Figure~\ref{9panela} show probability contours (solid lines)
depicting the likelihood, $P_a$, that a given combination of
$\alpha$ and $\beta$ reproduces the distribution of
velocities in the data for a minimum mass ratio of 
$q=0.02$ and binary fraction of $F=1.0$.  The 3x3 grid
corresponds to $r_{in}=0.020~AU$ (left column), $r_{in}=0.063~AU$ 
(middle column), $r_{in}=0.200~AU$ (right column), and
$r_{out}=10000~AU$ (upper row), $r_{out}=1000~AU$ (middle
row), $r_{out}=100~AU$ (lower row).  The horizontal line in
each panel marks the nominal $\beta=0$ (\"Opik's Law)
distribution of separations.  The dotted contours in each panel
depict the predicted fraction of type Ib/c supernovae 
in our Monte Carlo models.  These contours run from 10\% to 40\% 
in increments of 5\%.

The panels in Figure~\ref{9panela} demonstrate the
sensitivity of the best-fitting solutions for $\alpha$ and
$\beta$ to the adopted $r_{in}$ and $r_{out}$.  Each panel
contains a crescent-shaped ridge of highest likelihood
contours ($P_a\geq50\%$).  Values outside of this
region drop to probabilities $<10$\% over much of the
plotted parameter space. These contours show that the
best-fitting $\alpha$ and $\beta$ are correlated---larger
orbital separations are required (i.e., larger $\beta$) when
mass ratios are peaked toward unity, while smaller orbital
separations are required for mass ratio distributions with
secondaries drawn from a field star IMF.  The ridge of peak
likelihood separates parameter space into two regimes: to
the left of the ridge of solid contours the models fail to
produce enough high-velocity systems and are inconsistent
with the data; to the right of the peak contour ridge, the
models produce too many high-velocity systems relative to
the data.

For a flat distribution of orbital separations described by
\"Opik's law, the best-fitting values for $\alpha$ lie in the
broad range $-1.2<\alpha<1.0$ depending on the choice of
radial limits.  For the most reasonable choices
$r_{in}=$0.063--0.200~$AU$ and $r_{out}=1000~AU$, $\alpha$
lies in the range -0.6 -- 0.0, broadly consistent with
observational estimates of
\citep{kouwenhoven05,kouwenhoventhesis,garmany}.  The dashed
contours depicting $f_{Ib/c}=$0.25--0.35 intersect the
$\beta=0$ line and the solid contour ridge, indicating a
Ib/c fraction about twice the older observational estimates
of 15\% \citep{mannucci, cappellaro99} but consistent with
the most recent determinations of 30\%$\pm$11\% \citep{li,
leaman}.  Ib/c fractions of $f_{Ib/c}<$0.20 are inconsistent
with the data in all panels except the upper-right which
shows the (perhaps unrealistically) large radial limits of
$r_{in}=0.20~AU$ and $r_{out}=10,000~AU$.  It is noteworthy
that, in all of the panels, mass ratio distributions
described by $\alpha<-2$ are unlikely (contours levels
$<$30\%) even if the distribution of orbital separations
favors close companions (small $\beta$). Furthermore, any
$<\alpha<-1$ for the most plausible radial limits in the
middle or upper middle panels would require $f_{Ib/c}>0.40$
and $\beta<-0.3$, in disagreement with observational
results.

\begin{figure}
\epsscale{1.0}
\plottwo{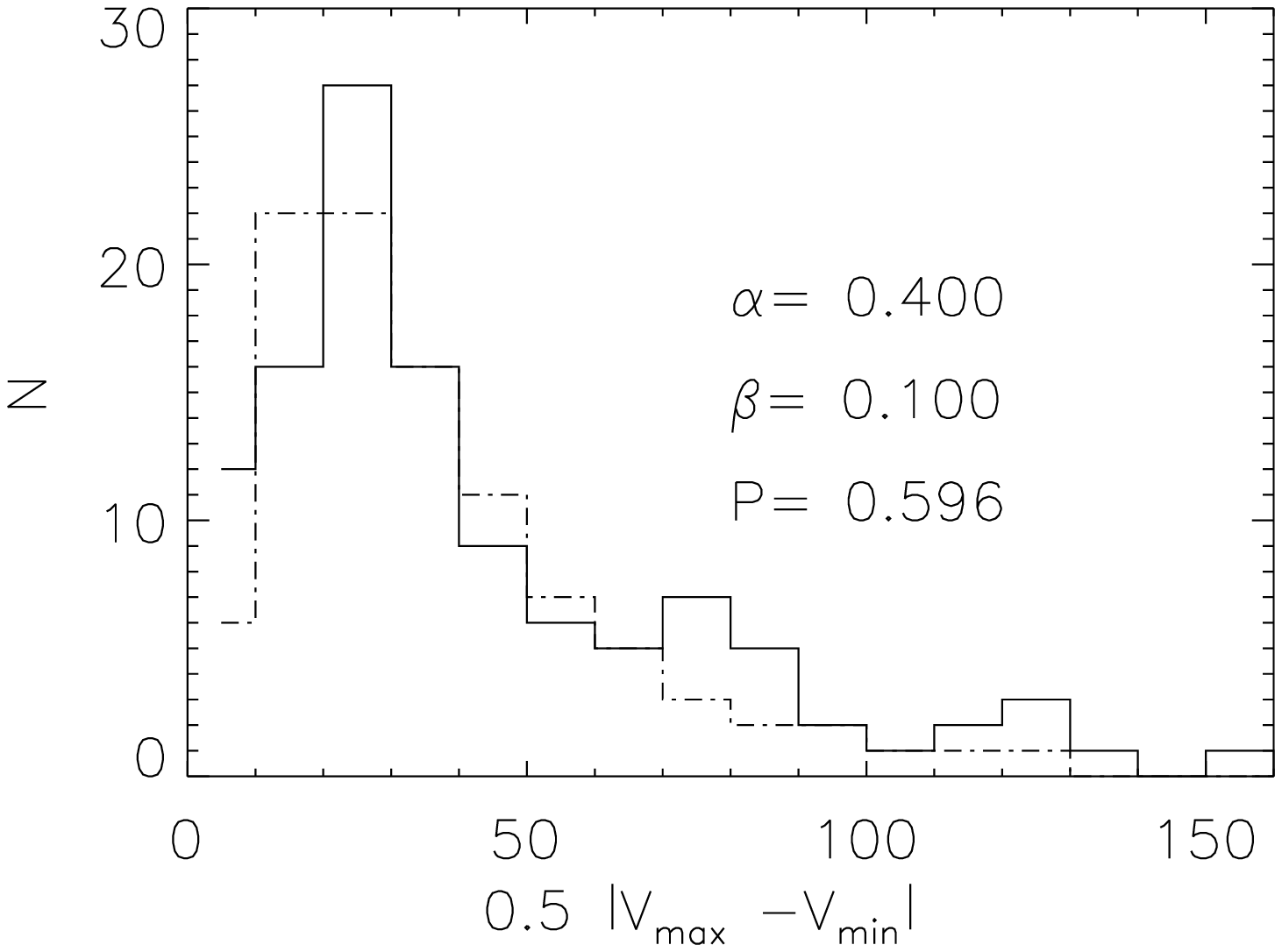}{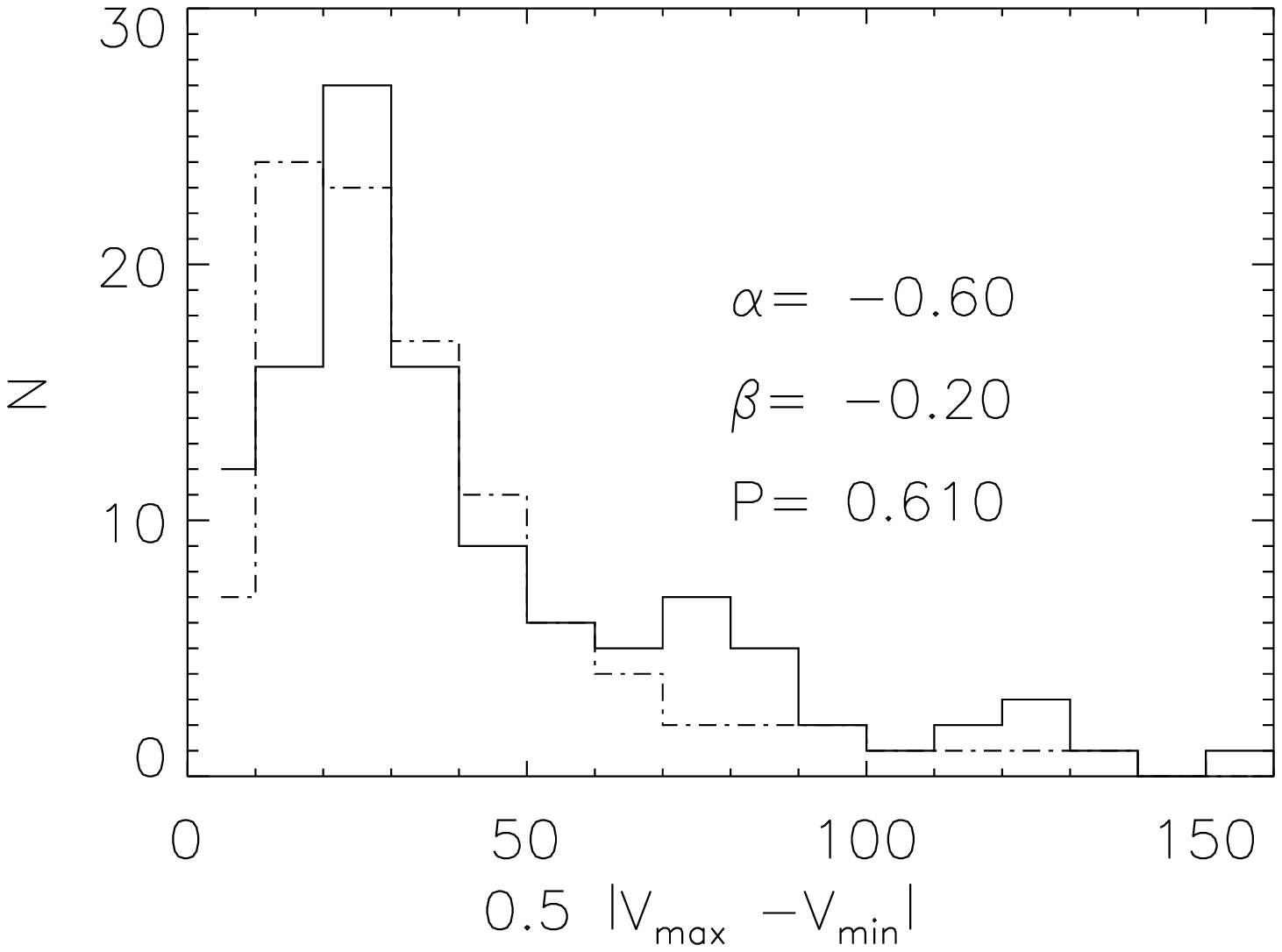}
\caption{Histograms showing the distribution of velocity semi-amplitudes,
$V_h$, for the data (solid line) and a particular set of
Monte Carlo simulations with the 
indicated values of $\alpha$ and $\beta$ (dashed line).  Each panel is 
labeled with the probability, $P(V_{h})$, that the two histograms are drawn 
from the same parent
population.  Both combinations of $\alpha$, $\beta$ yield probabilities
of $P_a\sim0.60$ and lie along the ridge line of peak likelihood from
the middle-right panel in Figure~\ref{9panela}. 
Either values of $\beta\simeq0.10$ paired with $\alpha\simeq0.4$
or  $\beta\simeq-0.20$ paired with $\alpha\simeq-0.6$ provide
similarly good matches to the data. 
    \label{hist9p}}
\end{figure}

Figure~\ref{hist9p} shows the distribution of observed and
simulated velocity semi-amplitudes, $V_h$, (solid and dashed
lines, respectively) for a subset of models that produce
$P_a\simeq60\%$ from the contour ridge in the middle-right
panel of Figure~\ref{9panela}.  The left panel of
Figure~\ref{hist9p} shows a simulation with $\alpha=0.4$,
$\beta=0.10$, and the right panel shows $\alpha=-0.6$,
$\beta=-0.20$.  Each panel is labeled with the probability
that the two histograms are drawn from the same parent
population.  Both combinations of $\alpha$, $\beta$ yield
probabilities of $P(V_h)\simeq0.60$ and lie along the
ridge line of peak likelihood.
There is good agreement between the data and simulations
 based on the two-sided K-S ``D''
statistic which quantifies the maximum difference between the
cumulative distribution functions of the two data sets.  An
analysis of the cumulative distribution functions shows that,
for these best-fitting models, the maximum differences occur
in the range 10--40 \kms\ and 80--100 \kms\ in roughly equal
proportion.  This means that, in some Monte Carlo
iterations, differences in the high-amplitude tail of the
velocity distributions limit the agreement between the
models and data; in other iterations, the models fail to
produce the observed velocity distribution on the
low-amplitude side of the histogram.  The lack of systematic
trends in the velocity at which the K-S ``D'' maximum occurs
may be taken as an indication that the comparison is
limited by sample size and measurement uncertainties.  If, for
example, the K-S ``D'' maximum difference always occurred at
small $V_h$ amplitudes, we would suspect a systematic
problem with the adopted velocity uncertainties which are
responsible for the objects in the lowest velocity bins.
If, on the other hand, the K-S ``D'' maximum difference
always occurred at large $V_h$ amplitudes, we might suspect
the presence of a second population of close binaries which
was responsible for the large-amplitude systems.  We will
return to this latter possibility in a subsequent section.
   
\begin{figure}
\plotone{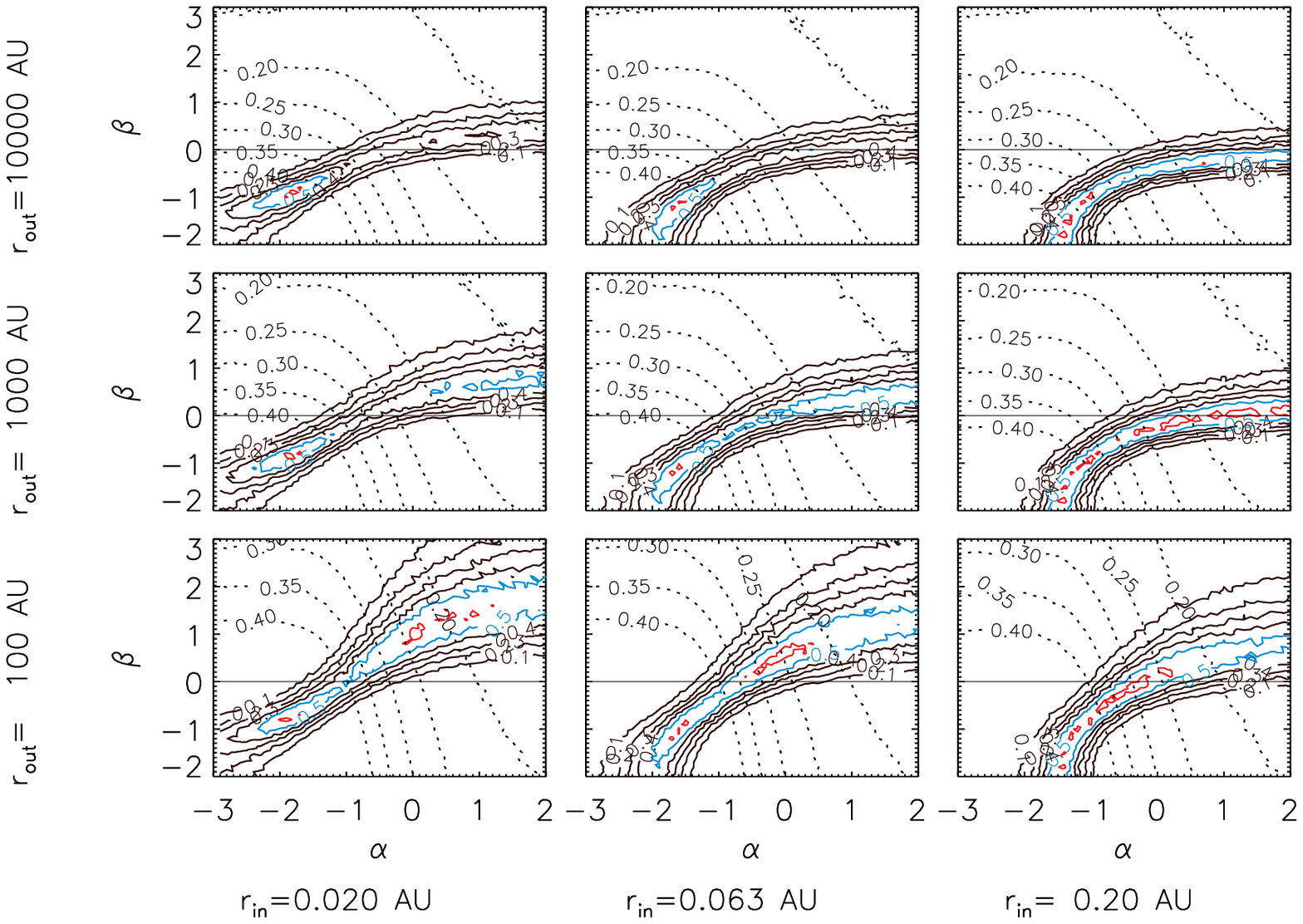}
\caption{Probability contours as in Figure~\ref{9panela}
except for a binary fraction of $F=0.8$.
 \label{9panelb}}
\end{figure}

Figure~\ref{9panelb} shows contour plot comparisons of the
data with Monte Carlo simulations in the same manner as
Figure~\ref{9panela}, except for a binary fraction of
$F=0.8$. The overall level of agreement between the
models and the data is lower in each panel by $\sim$10\% when compared
with the corresponding panel of Figure~\ref{9panela}.
 The contours are qualitatively similar to those
in Figure~\ref{9panela} in that the ridge of peak likelihood
reaches $P_a\sim50\%$ in most panels, but it is shifted slightly
toward larger $\alpha$ and smaller $\beta$.  Variations
between panels once again highlight the sensitivity of the
best-fitting solutions to the adopted radial boundary
conditions.  For instance, the middle-right panel
($r_{in}=0.20$, $r_{out}=1000$) shows a peak likelihood of
$P_a>60\%$ for an \"Opik's Law distribution of $r$, but only
for $\alpha>0.5$, meaning that mass ratios must be peaked
toward unity.  This is because, with fewer binaries, the
observed levels of velocity variation can only be reproduced
with larger mass ratios and/or smaller separations.  Thus,
the \citet{garmany} mass ratio distribution peaked toward
unity can be consistent with the data, but only if the
binary fraction is $F<0.8$ and only for somewhat large values of
$r_{in}=0.20~AU$, i.e., larger than many of the systems
observed by \citet{garmany}.  For the most plausible radial limits 
($r_{in}=0.063$, $r_{out}=1000$) in the center panel, the
most probable value for $\alpha$ is $\sim0.0$
for $\beta=0$. This implies $f_{Ib/c}=$0.20--0.25,
at the lower end of current observational constraints
\citep{li, leaman}.  Any of the other panels for plausible radial limits
(upper-middle or middle-right panels) require
$f_{Ib/c}<$0.20, a value that is increasingly inconsistent with
observational limits.
     
\begin{figure}
\plotone{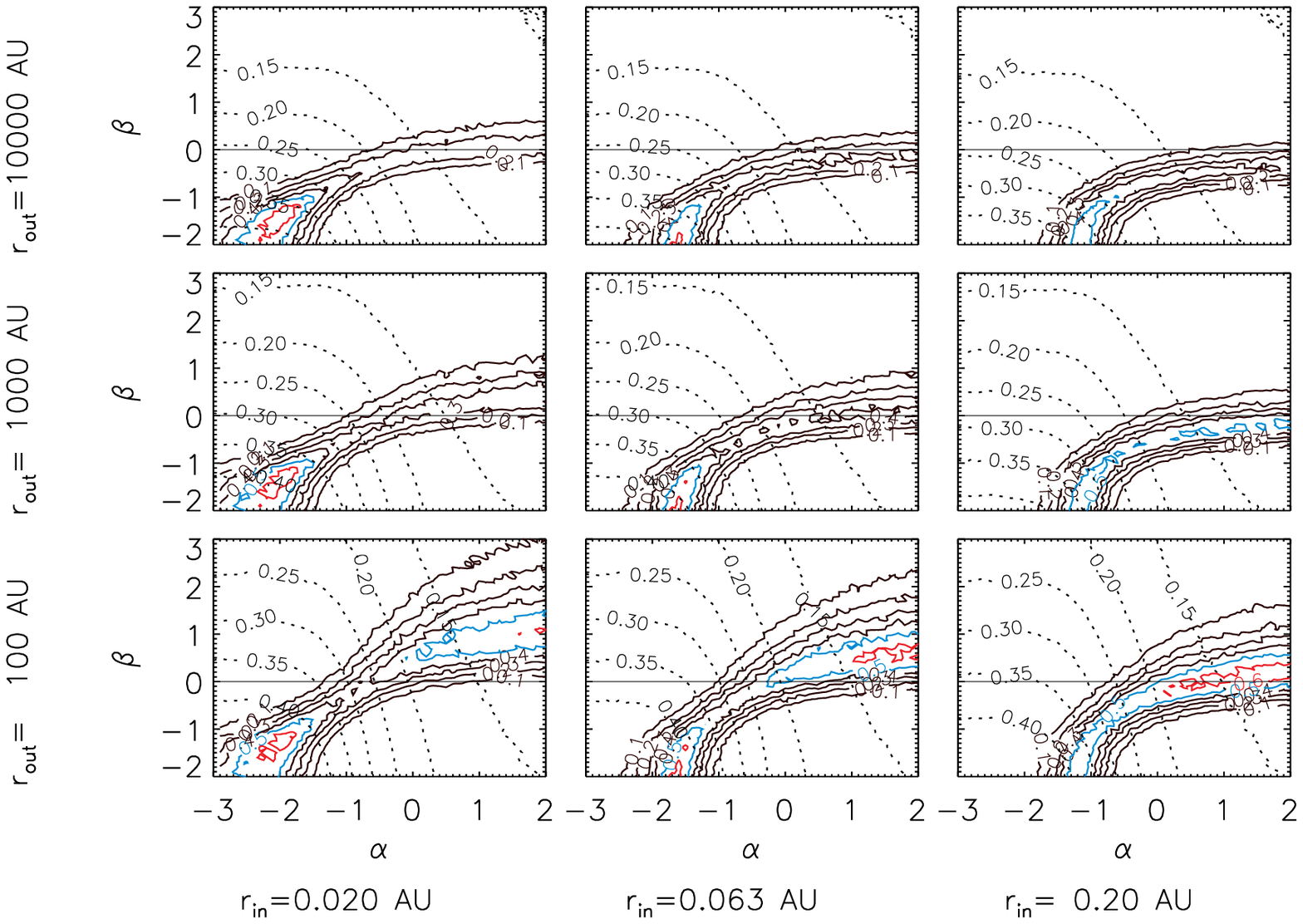}
\caption{Probability contours as in Figure~\ref{9panela}
except for a binary fraction of $F=0.6$.
 \label{9panelc}}
\end{figure}

Figure~\ref{9panelc} shows contour plot comparisons of the
data with Monte Carlo simulations in the same manner as
Figure~\ref{9panela} for binary fractions of $F=0.6$. 
Compared with the corresponding panels of Figure~\ref{9panelb},
the solid contours in each panel shows lower levels of agreement
between models and data by about 10\%.  This is especially true in the
middle and upper-middle panels which represent the most plausible 
radial boundary limits ($r_{in}=0.063-0.200$, $r_{out}=1000-10000$).
 These panels also show that mass ratio distributions with $\alpha<-2$
are highly inconsistent with the data even if the the orbital
separations are preferentially very small ($\beta<-1$) and
$r_{in}=0.020$ $AU$.  Thereby, the field IMF distribution for
secondary star masses is highly improbable given that
$r_{in}=0.020$ would require many systems in common
envelopes even for main-sequence primaries.  Reasonable
agreement with the data ($P_a>50\%$) is only achievable with
$\alpha>0$ and with $r_{out}=100$ $AU$ (lower row).
Because $r_{out}=100$ $AU$ is implausible (no long-period
companions!) and any of the panels for the most probable
radial limits (middle or upper-middle) require
$f_{Ib/c}<$0.15, at odds with current estimates \citep{li,
leaman}, we conclude that Figure~\ref{9panelc} contains no
acceptable portion of parameter space where a good fit
between the models and data can be achieved.  A binary
fraction as of $F=0.6$ or lower is highly improbable in Cyg
OB2.

In summary, the sequence of Figures~\ref{9panela},
\ref{9panelb}, \ref{9panelc} shows the degeneracies between
$\alpha$ and $\beta$ and the sensitivity of the best-fitting
solutions to the adopted boundaries of the power-law
parameterization of separations, $f(r)$.  Distributions of
$q$ favoring extreme mass ratios $\alpha<-2$ universally
require very small separations ($\beta<-1$) in disagreement
with observations supporting the canonical flat \"Opik's Law
distribution.  For plausible inner and outer boundaries, the
most probable values for $\alpha$ lies in the range $-0.6$
-- $0.0$ for $F=1.0$ and $0.0$ -- $1.0$ for $F=0.8$.  Binary
fraction $F<0.8$ would imply low fractions of type Ib/c
supernovae, $f_{Ib/c}<$0.20, inconsistent with current
estimates.

\subsection{Comparison with the Hogeveen Distribution of $q$}

\citet{hogeveen90,hogeveen92} proposed that $f(q)\propto q^{-2.0~-~-2.7}$ for
$q>q_0$ where $q_0=0.3$ for B star primaries, and
$f(q)\propto q^0$ (i.e., flat) for $q<q_0$. The relative
numbers of systems above and below $q_0$ is not given.  We
used an adapted Monte Carlo code to perform comparisons
between the data and the predictions of the
\citet{hogeveen92} mass ratio distribution.  
We adopted a two-component power-law distribution of $q$
with variable normalization for each component so that they
yield the same $f(q)$ at $q_0$.  The component with $q<q_0$
has fixed $\alpha(q<q_0)=0$ while $\alpha(q>q_0)$ is allowed to vary
freely.  This effectively
means that the fraction of systems with $q>q_0$ varies with
$\alpha$.  For $\alpha=2.0$, 90\% of systems have $q>q_0$;
for $\alpha=0.0$, 70\% of systems have $q>q_0$; for
$\alpha=-2.0$, 48\% of systems have $q>q_0$.  We compared
the velocity distribution of the data to the expectations of
Monte Carlo simulations in the same manner as for the simple
power-law parameterizations of $f(q)$ and $f(r)$ above.  We
note that in the limiting case $\alpha(q>q_0)=0$, the predictions
of the \citet{hogeveen92} mass ratio distribution and the 
power-law formulation discussed above are identical, within
the noise of the Monte Carlo procedure employed.

\begin{figure}
\plotone{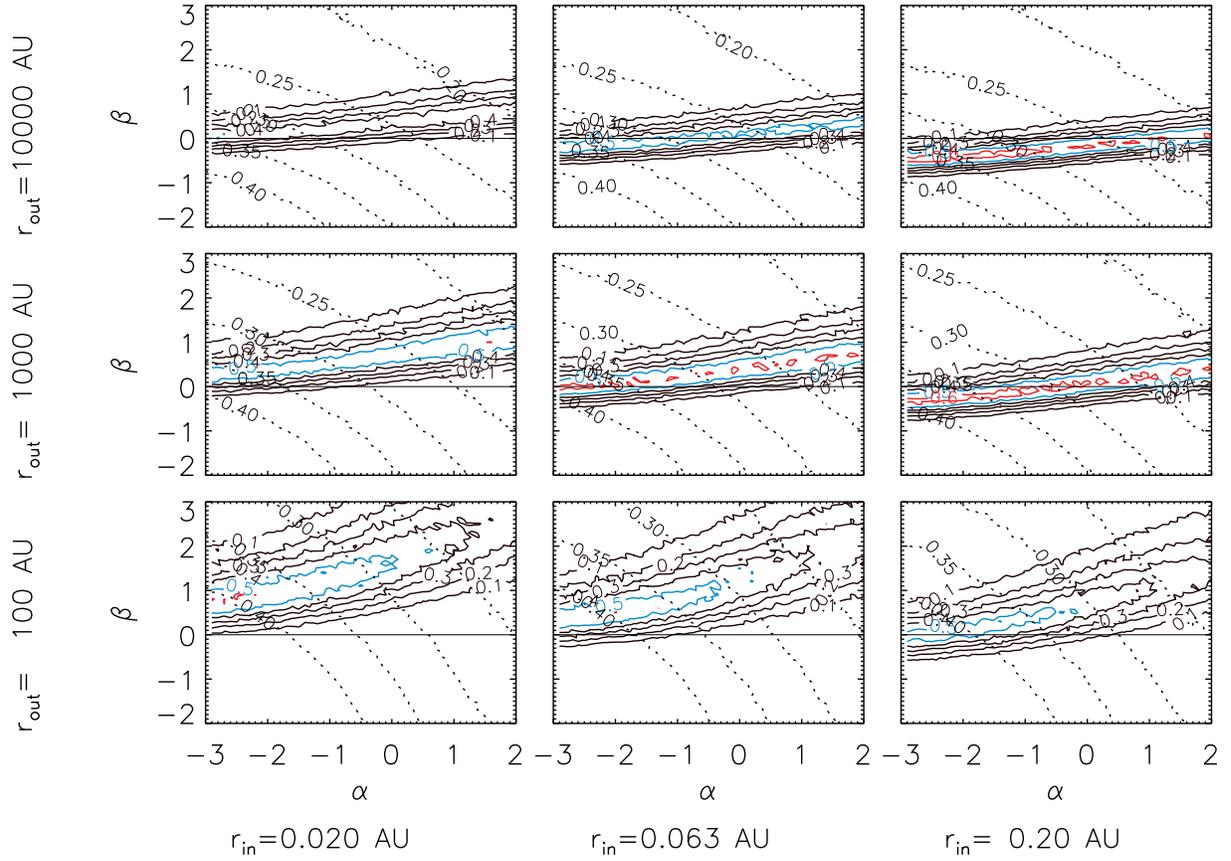}
\caption{Probability contours from 
Monte Carlo simulations similar to Figure~\ref{9panela}
but for the mass ratio distribution given by \citet{hogeveen92}.
  \label{9panelH}}
\end{figure}

Figure~\ref{9panelH} shows probability contour plots as a
function of $\alpha(q>q_0)$ and $\beta$ for a range of $r_{in}$ and
$r_{out}$ as in Figure~\ref{9panela}.  The ridge of peak
likelihood exceeds 60\% in some panels.  In particular, the
center ($r_{in}=0.063$) and middle-right ($r_{in}=0.20$)
panels show best agreement with the data.  Both panels have
$r_{out}=1000$ $AU$.  The center panel shows that for the
nominal \"Opik's Law distribution of separations, the most
probable values for $\alpha$ lie in the range
$-3.0<\alpha<-2.0$, consistent with the Hogeveen formulation
for $q>q_0$.  However, the predicted fraction of
type Ib/c supernovae is $f_{Ib/c}=$0.35--0.40, at the upper edge of
current estimates \citep{li, leaman}.  
In most other panels, $\alpha<-2$ is inconsistent with the data.
The middle-right panel shows that for the
\"Opik's Law distribution of separations, the most probable
values for $\alpha$ lie in the range $-1.0<\alpha<0.0$.
Hence, the most probable value for $\alpha$ for the Hogeveen
mass ratio distribution is much more sensitive to the choice
of $r_{in}$ than for the simple power-law distribution of
$q$.  This is because the peak contour ridge in each panel
is nearly horizontal, a consequence of the fact that the
actual distribution of mass ratios over the whole range
$q=$0.02 -- 1 varies comparatively little because the
portion with $q<q_0$ is unchanging. Under the Hogeveen
formulation, the two-component power law with
$\alpha(q<q_0)=0$ and $\alpha(q>q_0)=-2$ essentially mimics
a single power-law of intermediate slope, $\alpha\simeq
-0.6$.  The predicted fractions of type Ib/c supernovae in
the middle and middle-right panels are $f_{Ib/c}=$0.35--0.40
and $f_{Ib/c}\simeq$0.30 respectively. 

\begin{figure}
\plotone{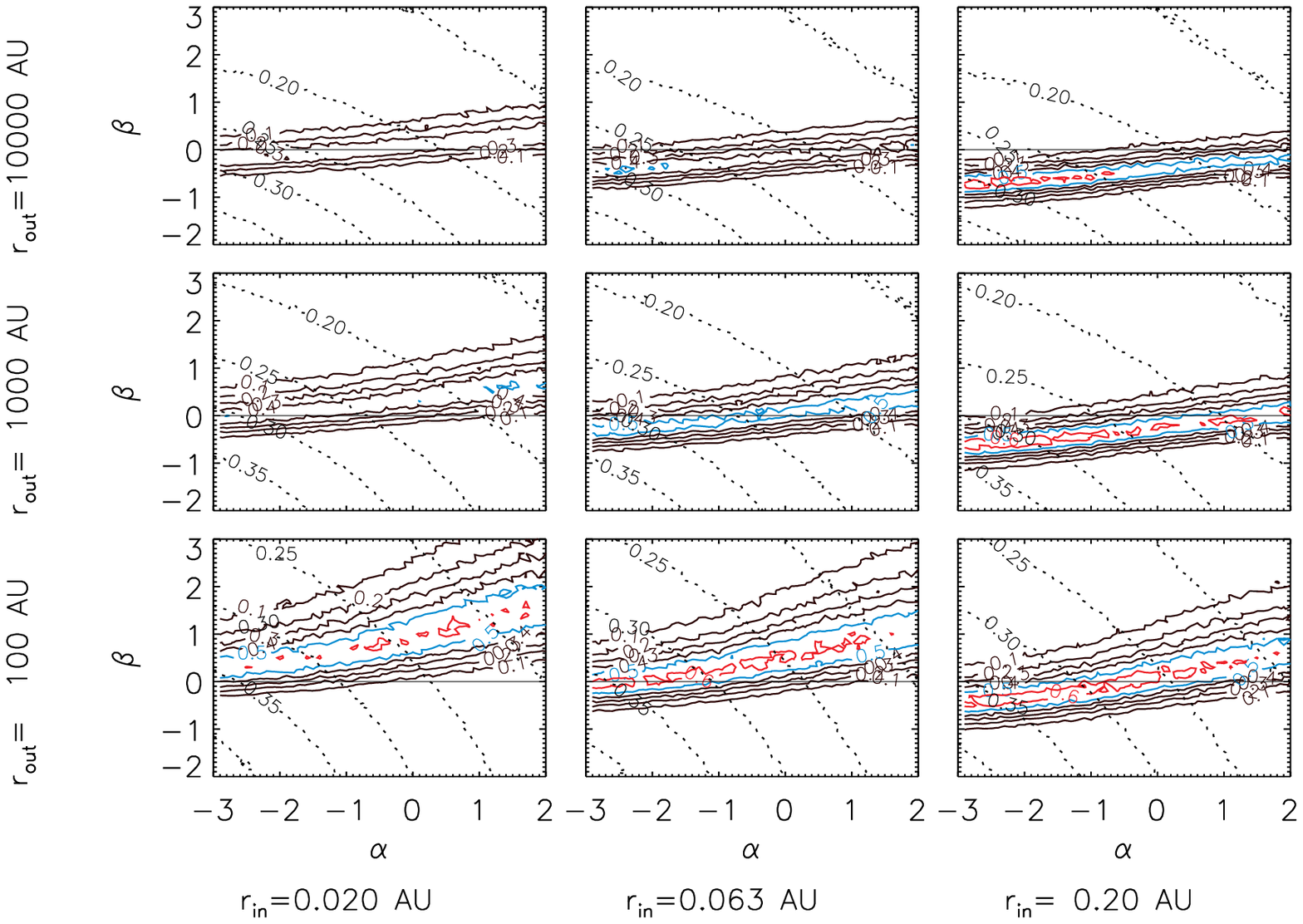}
\caption{Probability contours from 
Monte Carlo simulations similar to Figure~\ref{9panelH}
but for a binary fraction of $F=0.8$.
  \label{9panelH8}}
\end{figure}

Figure~\ref{9panelH8} shows probability contour plots as a
function of $\alpha$ and $\beta$ for a range of $r_{in}$ and
$r_{out}$ as in Figure~\ref{9panelH}, except for a binary
fraction of $F=0.8$.  The overall level of agreement between
the models and the data is lower in each panel by $\sim10\%$
when compared with the corresponding panel of
Figure~\ref{9panelH}.  The contours of peak likelihood are
shifted toward larger $\alpha$ and smaller $\beta$ relative
to Figure~\ref{9panelH}.  In the center panel, the peak
probabilities exceed 50\%, and for an \"Opik's Law
distribution of separations, the best fitting value of
$\alpha$ lies in the range $-1<\alpha<0.3$, inconsistent
with the \citet{hogeveen92} value of $\alpha=-2$.   In the
middle-right panel, the peak probabilities exceed 60\%, but
\"Opik's Law would require $\alpha>0.8$.  The predicted
fraction of type Ib/c supernovae is $f_{Ib/c}=$0.20-0.25,
at the lower end of current estimates.  The lower-middle
panel shows probabilities exceeding 60\% for \"Opik's Law
and $\alpha\simeq-2$.  However, this requires $r_{out}=100$
$AU$, and we argue previously that such a small outer bound
on companion separations is not supported by observations.
In summary, the \citet{hogeveen92} mass ratios distribution
and a binary fraction of $F=0.8$ produces slightly poorer
overall agreement with the data.  The best fitting models
consistent with \"Opik's Law require $-1<\alpha<1$ and
$r_{in}=0.063$--$0.200$ and $r_{out}\sim1000$ $AU$.
Simulations with $F\leq0.6$ show comparatively poor
agreement with the data, similar to Figure~\ref{9panelc},
and would require $f_{Ib/c}<$0.15, a value inconsistent with
current measurements.

\subsection{Comparison with a ``Twin'' Component Model Distribution of $q$ }

Given ample evidence in the literature for a significant
``twin'' population among short-period binaries
\citep{ps, tokovinin, lucy, lucy06}, we used a modified Monte Carlo
code to simulate the expected velocity amplitudes from a
two-component $q$ distribution.  The dominant component has
a power-law distribution, $f(q)\propto q^\alpha$ and
$f(\log{r})\propto [\log(r/r_{in})]^\beta$ between the limits
$r_{in}=0.063$ $AU$ and $r_{out}=1000$ $AU$, with $\alpha$
and $\beta$ being free parameters, as described previously.
The twin component comprises a fraction, $T_{frac}$, of the
systems.  We explore a range of twin fractions,
$T_{frac}=$0.40, 0.25, and 0.10. Allowing for the
possibility that $F$ may be less than unity, we define
$T_{frac}$ to be the fraction of systems with $q\simeq1$
{\it among those that are binary systems}.  Motivated by
observations that ``twin'' systems have small separations,
presumably because they form within a causally connected
region, we adopt a log-normal distribution with mean
separation, $<r_{Twin}>$, and Gaussian width,
$\sigma_{\log{r}}$.  We explore a range of typical
separations, $<r_{Twin}>=$1.0, 3.9, and 15 $AU$ (i.e.,
$<\log{r_{Twin}}>=$0.0, 0.6, and 1.2 ).  The smallest of
these mean separations, $<r_{Twin}>=1$ $AU$, is
representative of the orbital distances required to produce
OB star twins with orbital periods in the range few--60 days
like the samples used by \citep{ps,tokovinin,lucy} to infer
the presence of a twin population.  We adopt
$\sigma_{\log{r}}=0.6$ for the width of the twin
population's orbital separation distribution. (Exploration
of larger and smaller values for $\sigma_{\log{r}}$
yielded poorer levels of agreement with the data and we do
not show these results in the interest of brevity.)  The
mass ratios for the twin component are uniformly distributed
between $q=$0.95--1.0.  The lower mass limit for the
power-law component described by $f(q)\propto q^\alpha$ is
$q_{low}=0.02$.

\begin{figure}
\plotone{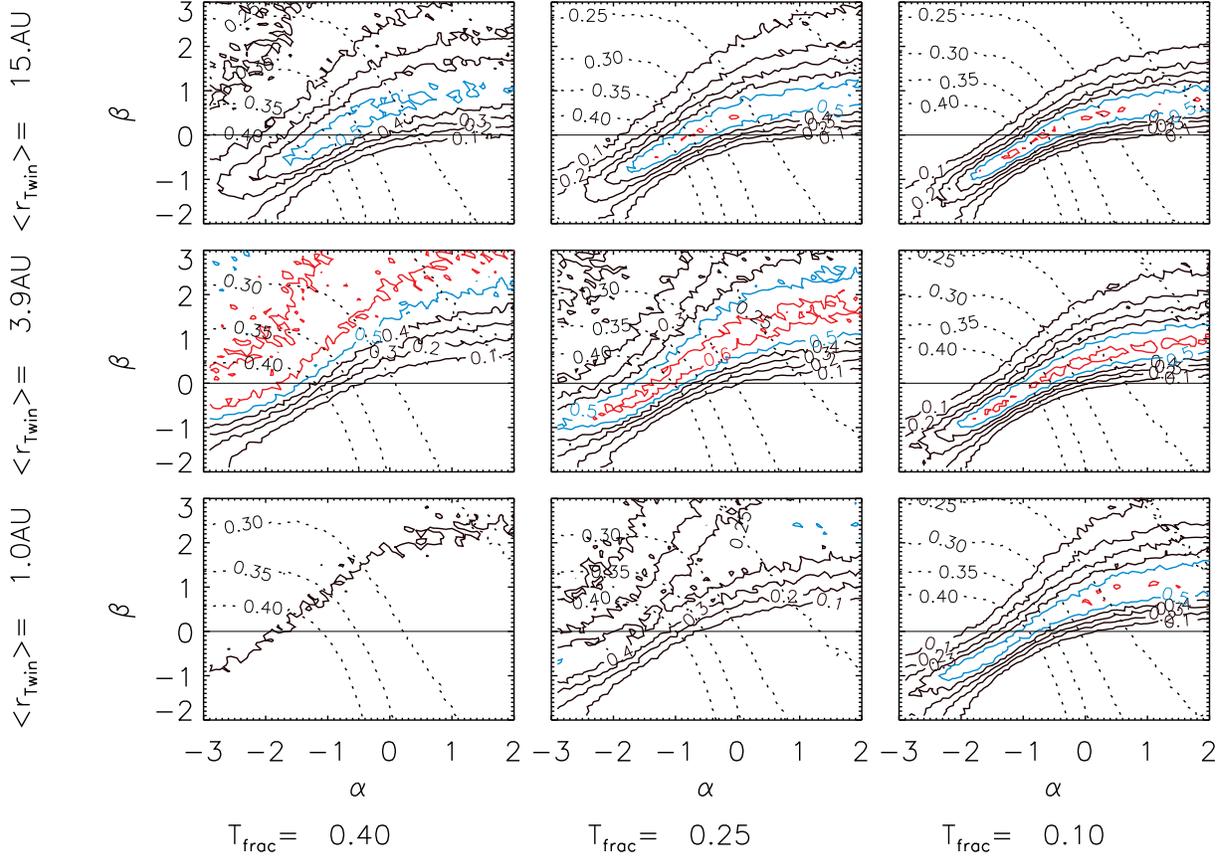}
\caption{Probability contours from 
Monte Carlo simulations similar to Figure~\ref{9panela}
but for a two-component mass ratio distribution.
This Figure shows a 3x3 grid of
contour probability plots for  $T_{frac}$=0.40 (left column), $T_{frac}$=0.25
 (middle column), $T_{frac}$=0.10 (right column), and
$<r_{Twin}>=15~AU$ (upper row), $<r_{Twin}>=3.9~AU$ (middle row),
$<r_{Twin}>=1.0~AU$ (lower row).
In all cases, the radial limits for the power-law distribution of the non-twin component
are $r_{in}=0.063$ $AU$ and  $r_{out}=1000$ $AU$.
  \label{9panelPS}}
\end{figure}

Figure~\ref{9panelPS} shows a 3x3 grid of contour
probability plots $T_{frac}$=0.40 (left column),
$T_{frac}$=0.25 (middle column), $T_{frac}$=0.10 (right
column), and $<r_{Twin}>=$15 $AU$ (upper row),
$<r_{Twin}>=$3.9 $AU$ (middle row), $<r_{Twin}>=$1.0 $AU$
(lower row).  The horizontal line in each panel denotes
\"Opik's Law ($\beta=0$).  The panels in the right column
having $T_{frac}=0.10$ closely resembles the center panel of
Figure~\ref{9panela} where no twin component is included,
indicating that twin fractions of $<$10\% do not measurably
alter the results.  In general, the lower row with
$<r_{Twin}>=$1.0 $AU$ shows relatively poor agreement with
the data---small mean separations yield too many
large-amplitude systems.  The peak probabilities exceed 60\%
in several panels, most significantly in the middle panel
corresponding to $<r_{Twin}>=$3.9~$AU$, $T_{frac}$=0.25 and
in the middle-left panel corresponding to
$<r_{Twin}>=$3.9~$AU$, $T_{frac}$=0.40.  In these panels
\"Opik's Law specifies best-fitting values for $\alpha$ in
the range $-3.0<\alpha<-1.3$. The predicted fractions of
type Ib/c supernovae are large in both cases, exceeding
0.40.  In the middle-left panel showing $<r_{Twin}>=$3.9
$AU$, $T_{frac}=0.40$, there is a broad region where the
peak probabilities exceed 60\% and reach 70\%.  In this
case, if \"Opik's Law obtains, then $\alpha<-2$, consistent
with the expectations of randomly pairing secondaries from
the field IMF with massive primaries.  Figure~\ref{9panelPS}
illustrates that two-component $q$ prescription of
\citet{ps} (45\% of systems with $q>0.95$, the remainder
with $\alpha=0$) is not consistent with the data unless
$\beta>0.5$.  That is, the probability of $\alpha=0$,
$\beta=0$ is $<30\%$ for all of the panels.  The presence of
any significant twin component appears to require
$\alpha\leq-1$.

\begin{figure}
\plotone{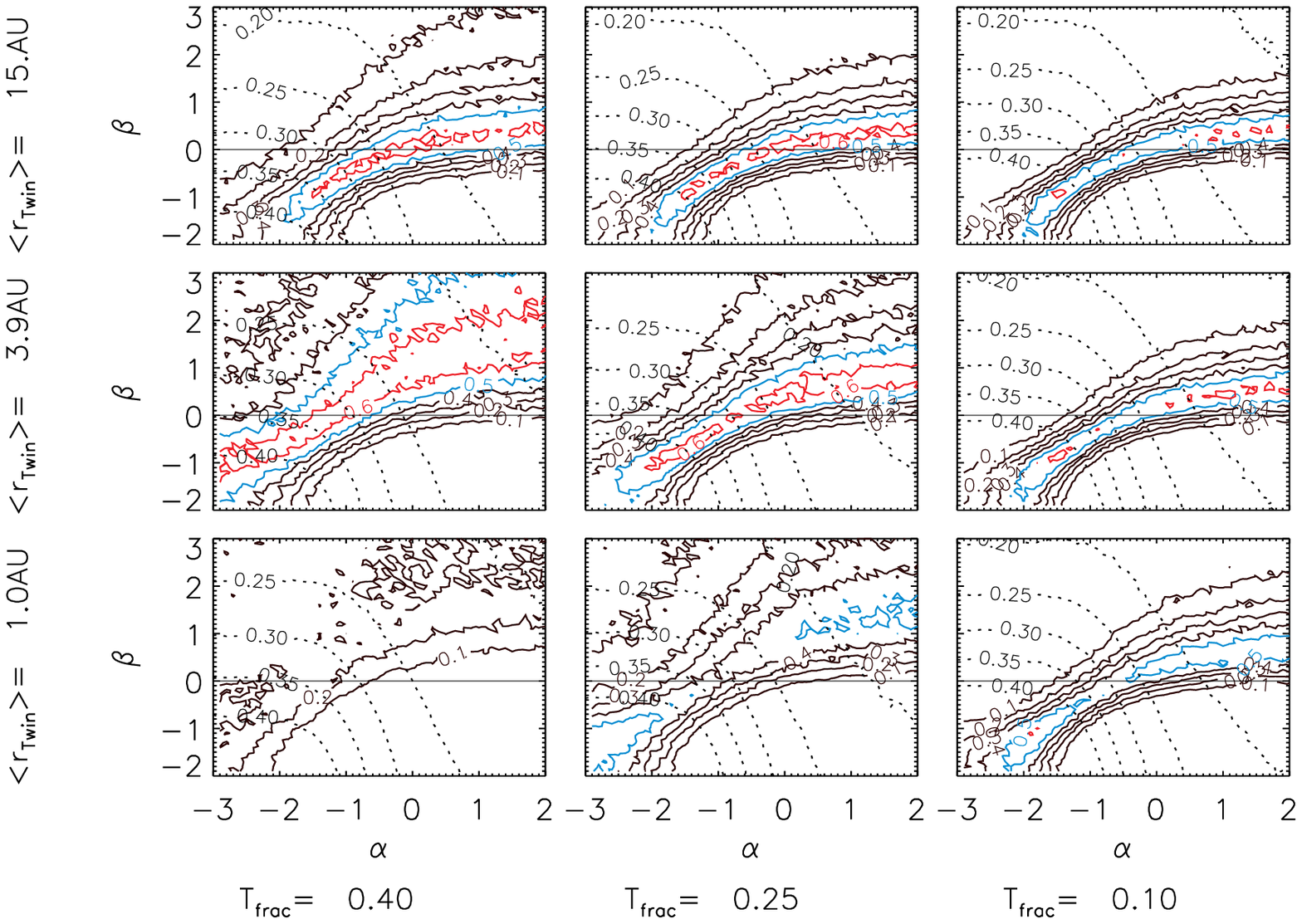}
\caption{Probability contours from 
Monte Carlo simulations similar to Figure~\ref{9panelPS}
but for a binary fraction $F=0.8$.
  \label{9panelPS8}}
\end{figure}

Figure~\ref{9panelPS8} shows a 3x3 grid of contour
probability plots similar to Figure~\ref{9panelPS}, but for
a binary fraction of $F=0.8$.  The peak contours once again
exceed 60\%, primarily in the upper two rows of panels.  The
ridge of peak likelihood is shirted toward larger $\alpha$
and smaller $\beta$ relative to Figure~\ref{9panelPS}.  In
most panels, the region with $P_a>0.60$ is smaller than in
the corresponding panels of Figure~\ref{9panelPS},
indicating slightly poorer overall agreement with the data.
For an \"Opik's Law distribution of separations, the most
probable values for $\alpha$ lie in the range -1.5 -- 0.0
for twin fractions $T_{frac}=$0.25--0.40 and
$<r_{Twin}>=$3.9--15 $AU$.  Models with $<r_{Twin}>=$1 $AU$
produce generally poorer agreement with the data, having
peak likelihood $<50\%$.  The two-component $q$ prescription
of \citet{ps} (45\% of systems with $q>0.95$, the remainder
with $\alpha=0$) is not consistent with the data unless
$<r_{Twin}>\simeq15$ $AU$.  At such large distances, OB star
twins would have much longer periods ($\sim10$ yr) than the
$\sim$30-day systems used to infer the presence of a twin
population \citep{ps, tokovinin, garmany}.  For $\beta=0$ 
and in panels with $P_a\geq0.60$
(middle, middle-left and upper-left), the predicted
fractions of type Ib/c supernovae are $f_{Ib/c}=$0.20--0.30,
in broad agreement with observations
\citep{li, leaman}. 

\begin{figure}
\plotone{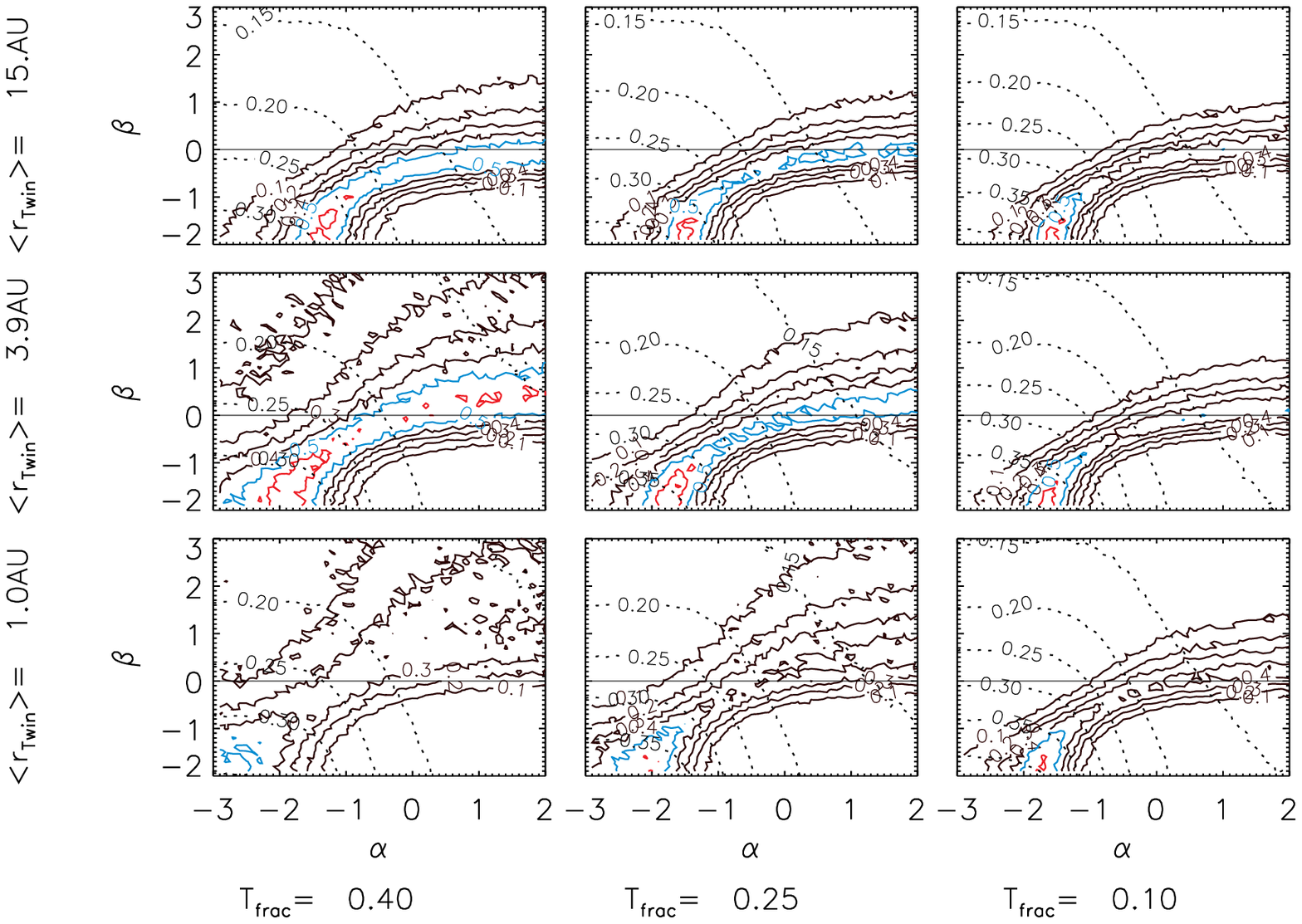}
\caption{Probability contours from 
Monte Carlo simulations similar to Figure~\ref{9panelPS}
but a binary fraction $F=0.6$.
  \label{9panelPS6}}
\end{figure}

Figure~\ref{9panelPS6} shows a 3x3 grid of contour
probability plots similar to Figure~\ref{9panelPS}, but for
a binary fraction of $F=0.6$.  The ridge of peak
likelihood is now shifted ever farther toward large $\alpha$,
and the peak probabilities do not much exceed 50\% in most
panels, indicating poorer agreement with the data compared
to Figure~\ref{9panelPS8}.  In those panels (middle, middle-left, 
upper-middle), \"Opik's
Law would require $\alpha>0$, in disagreement with
the \citet{ps} and field-star IMF distribution.  Furthermore, 
the type Ib/c supernova fraction is $f_{Ib/c}<$0.20,
in disagreement with observations.  Binary fractions 
$F\leq0.6$ appear improbable.

\subsection{Comparison with a Miller-Scalo Secondary Mass Distribution }

Power-law parameterizations of the stellar initial mass
function are appropriate only over a selected mass range.
More realistic descriptions of the IMF include multiple
power-law components with slopes that flatten toward lower
masses \citep[e.g.,][]{miller, scalo, kroupa93}.  We used a
modified Monte Carlo code to simulate the velocity
distribution expected if the secondary masses were drawn
from a distribution described by \citet{miller} as
parameterized in Equation 5 of \citet{eggleton}.

\begin{figure}
\plotone{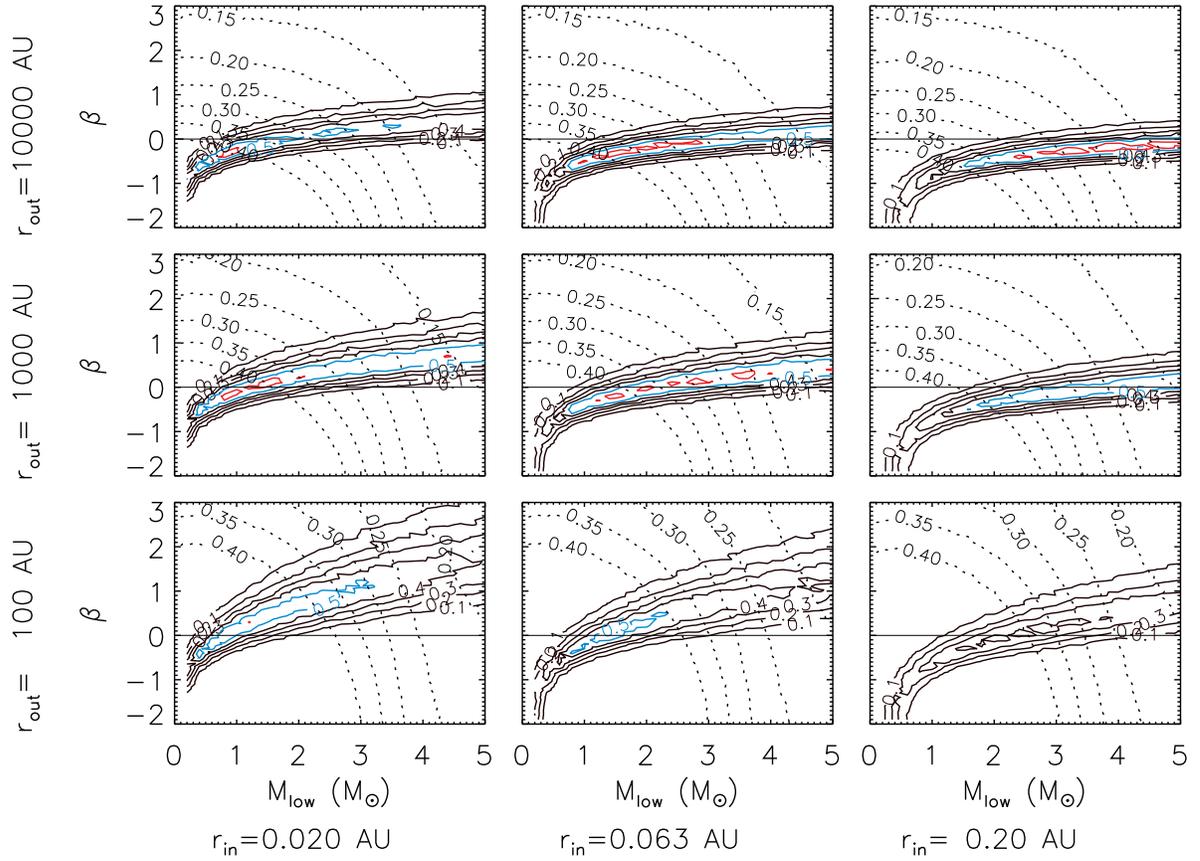}
\caption{Probability contours from 
Monte Carlo simulations similar to Figure~\ref{9panela}
but for a distribution of secondary masses described by 
the Miller-Scalo IMF with a lower mass cutoff shown 
on the ordinate of each panel.
  \label{9panelscalo}}
\end{figure}

Figure~\ref{9panelscalo} shows a 3x3 grid of contour
probability plots similar to Figure~\ref{9panela} except
that the ordinate in each panel is $M_{low}$, the lower mass
limit at which the \citet{miller} IMF is truncated. The
columns and rows of panels show a range of $r_{in}$ and
$r_{out}$, labeled in each panel, as in Figure~\ref{9panela}.
The peak probabilities marginally exceed 60\% in four
panels: middle-left, center, upper-middle, and upper-right.
Type Ib/c supernova fractions $f_{Ib/c}=$0.20 -- 0.40 are
broadly consistent with the data.  
However, in all four panels, the \"Opik's Law distribution of
separations shown by the solid horizontal line requires a
lower mass cutoff $M_{low}>1-3~M_\odot$. The probabilities
that the simulated data agree with the observations drop to
below 40\% at $M_{low}<1~M_\odot$ for $\beta<0$.  The
probabilities at low $M_{low}$ are somewhat higher in the
left column ($r_{in}=0.02$ $AU$), but we argue previously
that such small inner limits, coupled with such small $\beta$
are highly improbable physically.  Binary fractions $F<1$
would only exacerbate the problem, requiring even larger
lower mass cutoffs to the \citet{miller} IMF.  We conclude
that the data are grossly inconsistent with a \citet{miller}
type IMF extending below 2--4 $M_\odot$ for any plausible
distribution of separations.

\begin{figure}
\plotone{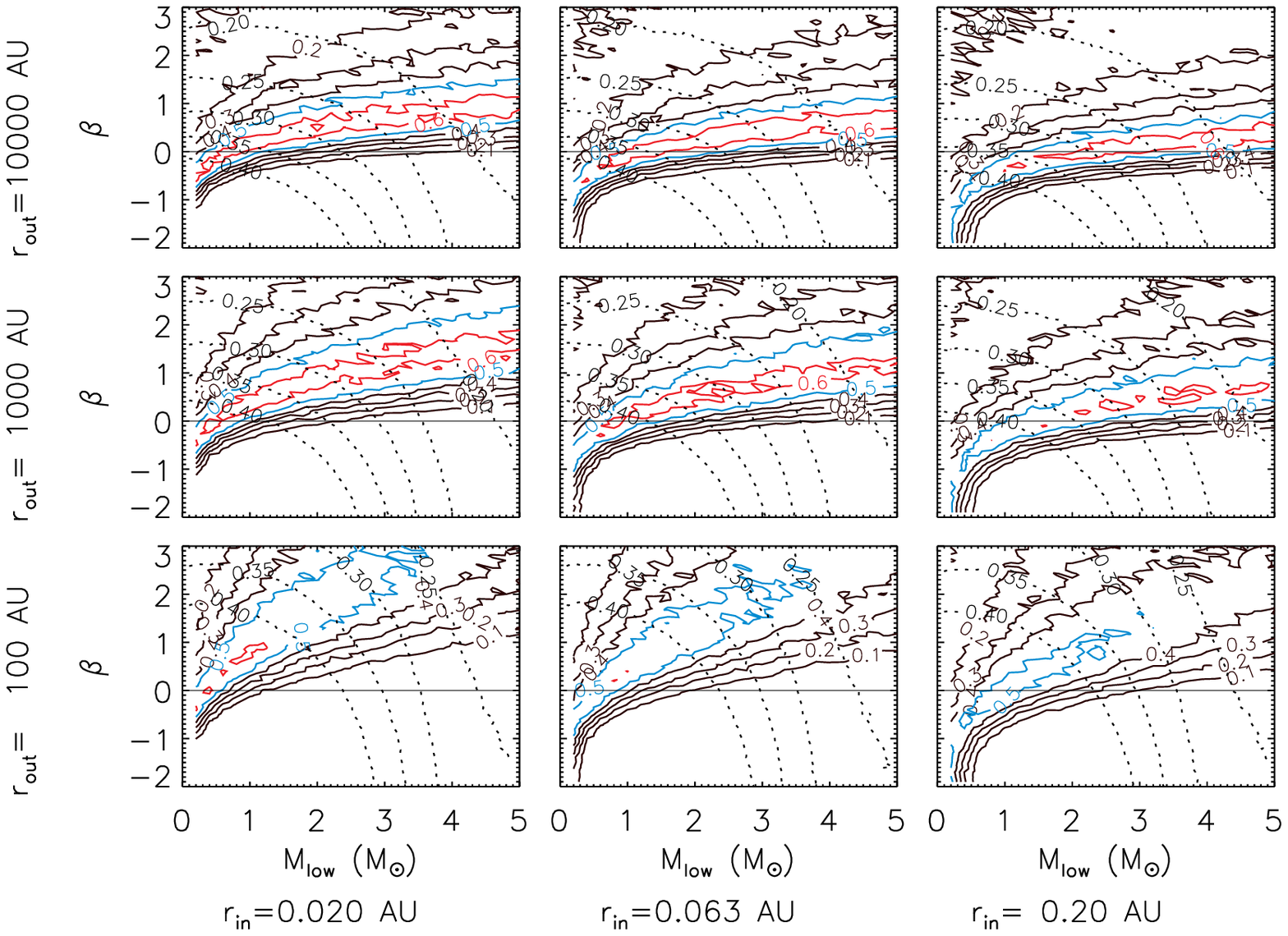}
\caption{Probability contours from 
Monte Carlo simulations similar to Figure~\ref{9panelscalo}
for a distribution of secondary masses described by 
the Miller-Scalo IMF but with a twin fraction
$T_{frac}=0.25$.  \label{9panelscalo25}}
\end{figure}

We tested whether a \citet{miller} type IMF might be made
more consistent with the data by the addition of a ``twin''
component in the same manner described previously.
Figure~\ref{9panelscalo25} shows a 3x3 grid of contour
probability plots similar to Figure~\ref{9panelscalo} except
that 25\% of the stars are now ``twins'' with $q=$0.95--1.0
with mean separations $<r_{Twin}>=3.9~AU$ and a log-normal
distribution having width $\sigma_{\log{r}}=0.6$.  The
highest contours peak in excess of 60\%, but these regions
lie above lower mass cutoffs of $M_{low}$=0.8 \msun\ for
plausible values of $r_{in}\geq0.063~AU$.  The \"Opik's Law
distribution of separations would require $M_{low}\geq0.5$
\msun\ where it intersects the peak contour ridge.  Even if
25\% of the companions are twins, a secondary star mass
function similar to the \citet{miller} IMF must have a lower
mass cutoff at relatively large masses in order to produce
the velocity distribution in the data.  The implied type
Ib/c supernova fractions are $f_{Ib/c}\geq$0.35 for the most
plausible radial boundary limits (middle and upper-middle
panels), at the upper end of currently measured ranges
\citep{li, leaman}.

\begin{figure}
\plotone{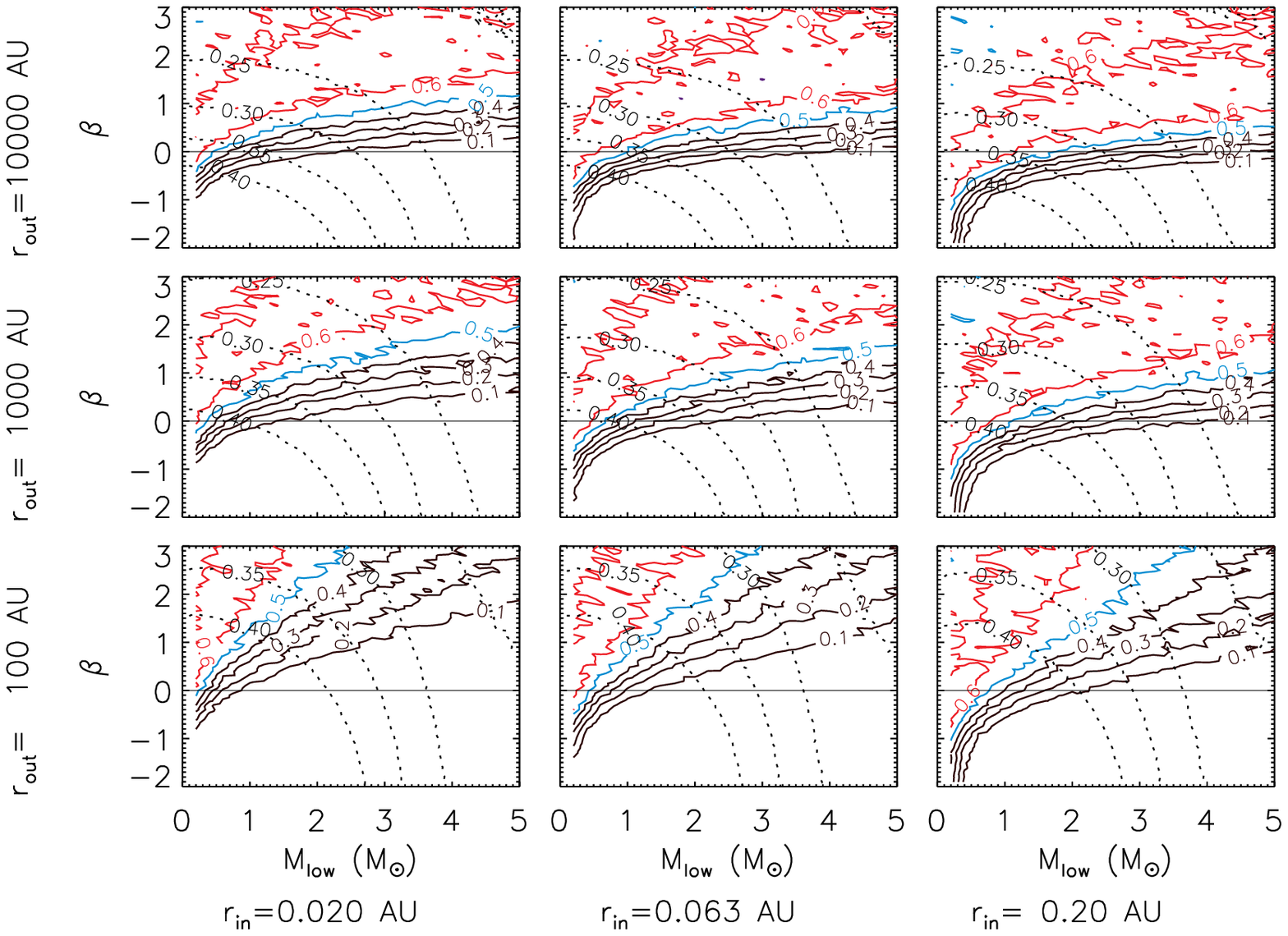}
\caption{Probability contours from 
Monte Carlo simulations similar to Figure~\ref{9panelscalo}
for a distribution of secondary masses described by 
the Miller-Scalo IMF but with a twin fraction
$T_{frac}=0.40$.  \label{9panelscalo40}}
\end{figure}

We tested the \cite{miller} type IMF against the data with a
``twin'' fraction of $T_{frac}=0.40$, similar to the
prescription of \cite{ps}.  Figure~\ref{9panelscalo40} shows
a 3x3 grid of contour probability plots similar to
Figure~\ref{9panelscalo25} for this series of Monte Carlo
comparisons.  The highest contours exceed 60\% over a wide
area and approach 70\%.  In nearly all panels, they extend
down to cutoff masses of $\sim0.1$ \msun.  In particular,
the four upper-right panels having $r_{in}\geq 0.063 ~AU$
and $r_{out}\geq 1000 ~AU$ show that the \"Opik's Law
distribution of separations intersects the ridge of peak
contours in the range 0.1 -- 0.3 \msun, i.e., in the range
where the \citet{miller} type IMF is flat, or turns over for
the lowest mass stars.  The implied type Ib/c supernova
fractions are $f_{Ib/c}\geq$0.35 for these panels, at the
upper end of, but consistent with, the currently measured
values \citep{li, leaman}.

\subsection{Limitations of the Modeling}

In order to compare the observations to the modeled velocity
distributions we have made a number of simplifying
assumptions.  Some of these assumptions may account for the
result that the best models agree with the data at only the
60--70\% level.  Nevertheless, we contend that none of these
assumptions affect the major conclusions.

We assume that all orbits are circular.  Although this is
unlikely to be the case for the majority of systems, it
provides a reasonable way to estimate the mean properties of
an ensemble of systems with a minimum of free parameters,
especially in the absence of well-sampled velocity curves
that would be required to measure eccentricities.  It is
also a reasonable approximation for close, massive binaries
where the orbits are likely to have circularized. Early-type
systems with periods less than several days are observed to
have eccentricities near zero \citep{giuricin} as are
systems with large fractional radii, $R_*/r>0.24$ where $r$
is the semi-major axis and $R_*$ the stellar radius
\citep{north, pan, zahn}. Low-mass binary systems have
eccentricities near zero for periods shorter than $\sim11$
days \citep{meibom}.  Without additional data to provide
secure periods for much for our sample, we can only say that
at least $\sim10$ systems appear to be in such short period
orbits.  Furthermore, although the velocity amplitude for a
primary star in an eccentric system is larger by a factor
$\sqrt{(1-e^2)}$ compared to a circular system of the same
semi-major axis, the rms velocity dispersion averaged over
the orbit is smaller by the same factor because the relative
radial velocity variations are small over most of the
orbital period.  It follows that the systematic effects of eccentric
orbits cancel to zeroth order and can be neglected
unless the majority of systems
are highly ecentric.

We assume that each primary star has a single companion that
dominates the observed primary kinematics.  In reality,
triple and quadruple systems probably exist among the
sample, but they are likely to be a small fraction of all
systems.  Among solar type stars the fraction of triple and
quadruple systems is $\leq4\%$ \citep{duquennoy}.  The
tertiary components, if any, are statistically likely to
have wide orbital separations and long periods, making their
observable dynamical influence minimal over few year time
baselines and at $\sim$10 \kms\ velocity precisions.

We assume that the measured velocity variations are due to
orbital dynamics of the primary star.  Stellar
photospheric line profile variations may be present among
some of the most massive stars, especially the evolved stars
in our sample (31 of \numstars\ or about 27\% are
post-main-sequence stars).  Line profile variations
attributed to atmospheric pulsations are observed in
$\geq$77\% of evolved O stars and in some Be stars
\citep{penrod, vogt} but rarely among dwarfs
\citep{fullerton96}.  These phenomena could mimic the
effects of bona fide orbital velocity variability in systems
observed with sufficiently high spectral resolution to see
these effects.  Neglecting this possibility would lead to an
overestimate of the binary fraction and bias the conclusions
in favor a larger $\alpha$ and/or smaller $\beta$.  Since
our sample consists predominantly of dwarfs and non-Be stars
(there are 3--5 Be stars), the impact of line profile
variations is minimal and is unlikely to dominate the
results.

Finally, we caution that the binary characteristics within a
single OB association may not be representative of the
binary properties in other massive starforming regions.  The
fraction of binaries, the distribution of mass ratios, the
initial orbital separations, and their eccentricities may
depend upon the {\it global} conditions in the molecular
cloud from which they formed.  \citet{garcia} highlight a
possible trend in the binary frequency with cluster density
and richness.  For example, the dense, massive molecular
clouds with densities $n_e >10^{7}$ $cm^{-3}$ which can
produce a super star cluster with $M>10^6$ $M_\odot$ within
a diameter of a few pc \citep{oconnell, elmegreen,
kobulnicky99, billett} may produce a different population of
binaries than those which generate the relatively diffuse OB
associations.  This would be especially true if binaries
result primarily from gravitational encounters between cloud
cores and/or massive protostars early in the star formation
episode.  On the other hand, it might be argued that the
production of binaries is primarily a {\it local} effect
driven by physics on the scale of an individual cloud core
($<10$ $AU$).  If massive stars form primarily through
mergers of intermediate mass stars in a dense core as
suggested by \citet{bonnell98}, then perhaps characteristics
of massive binaries are set by the masses and separations of
subclumps that fail to complete the merger process.
\citet{krumholz} provides a review of the predictions and
consequences of competing theories for massive star
formation.
  
\section{Conclusions}

We have analyzed the radial velocity data spanning 6 years
on a large sample of OB stars from the Cygnus OB2
Association to measure the probable system characteristics
among massive binaries.  We have compared the radial
velocity amplitudes to expectations from Monte Carlo models
based upon several popular formulations for the mass ratios
distributions in massive binaries.  These comparisons
highlight the allowed parameter space and the sensitivity of
the models to the key binary parameters: the true binary
fraction, $F$; the power-law slope $\alpha$ of the mass
ratio distirbution over the range $q_{low}<q<1.0$; and $\beta$,
the power law slope of the orbital
separations between the inner and outer radial boundaries, 
$\log(r_{in})$ and $\log(r_{out})$.  The data improve
upon prior studies by including a larger number of objects
in a common association over more observational epochs.  The
sample also consists of a young population with a (mostly)
uniform age of 2--3 Myr, thereby minimizing evolutionary and
selection effects inherent in many samples.

\begin{enumerate}

\item{Descriptions of the massive binary mass ratio
distribution and orbital separation distribution derived
from spectroscopic and direct imaging studies are biased by
selection effects and measurement limitations.  These
produce apparently conflicting prescriptions of massive
binary parameters.  Power-law parameterizations require
explicit boundaries over which the power law is valid.
These boundaries are often insufficiently well specified in
the published studies to allow meaningful comparisons given
that the conclusions are sensitive to the choice of $r_{in}$
and $r_{out}$, as shown in the Figures herein.  }

\item{Radial velocity amplitude data, such as we model here,
are consistent with a range of mass ratio distributions and
orbital radius distributions.  Although the models are
degenerate in $\alpha$ and $\beta$, the best fitting
power-law slopes are correlated.  Under the assumption that
the distribution of orbital separations is approximately
flat (\"Opik's Law $\equiv\beta=0$), the data allow only a
narrow range of $\alpha$ for given inner and outer radial
boundaries.  }

\item{Power-law parameterizations of $f(q)$ and $f(r)$ for
plausible boundary values $q_{min}=0.02$,
$0.063~AU<r_{in}<0.200~AU$, $r_{out}=10000~AU$, and $F=1.0$
are consistent with a mass ratio distribution described by
$-0.6<\alpha< 0.0$ assuming that $\beta\simeq0$, i.e., the
flat \"Opik's Law distribution of separations.  This is
consistent with $\alpha \simeq -0.4$ reported for adaptive
optics imaging studies of Sco OB2
\citep{kouwenhoven05,kouwenhoventhesis}.  Best-fitting Monte
Carlo Models are, on average, consistent with the data at
the $P_a=$60--70\% level.  Agreement between models and data is
limited by discrepancies between the cumulative distribution
functions of the model velocity amplitudes and the data both at
small velocities $V_h\sim20$ \kms\ and at large
velocities $V_h\sim80$ \kms\ in roughly equal proportions.
This result suggests that small sample size, measurement
uncertainties, and the possible presence of higher-order
systems in the data all act to preclude better agreement.  }

\item{Power-law models with binary fractions $F=0.8$ fit the
data slightly more poorly than models with $F=1.0$, and they
require larger mean mass ratios with $-0.4<\alpha<1.0$. }

\item{Power-law models with binary fractions $F\leq0.6$ fit
the data significantly less well, having $P_a<0.4$.  They
also require extreme values of $\alpha$, $\beta$, and
$f_{Ib/c}<0.20$ that are inconsistent with observational
studies. }

\item{The \citet{hogeveen92} distribution of mass ratios 
($\alpha=-2$ for systems with $q>0.3$ and $\alpha=0$ for
systems with $q<0.3$ ) for $r_{in}=0.063~AU$ and
$r_{out}=1000~AU$ can reproduce the data at the $\sim60\%$
level, comparable to the best-fitting power-law models with
$\alpha\simeq-0.6$.  This is because the two-component
\citet{hogeveen92} $q$ distribution effectively mimics a
power law with slope intermediate between $-2$ and $0.0$.
Given the ad hoc nature of the this formulation, its
additional parameters, and its inability to produce a
superior match to the data, we find no reason to prefer it
over a simpler power-law description for $f(q)$.  }

\item{Secondary star masses drawn from a \cite{miller} IMF 
are inconsistent with the data unless the IMF is truncated
at low masses below $\sim$2--4 \msun.   IMF truncation
at such high masses is required to produce the levels
of velocity variability seen in the data. {\it  The implication here is
that the physical conditions under which massive stars form 
are conducive to producing relatively close binaries with
correlated component masses. }    }

\item{Mass ratio distributions involving both a twin
component comprising 25--40\% of multiple systems and a
power-law component with $f(q)\propto q^0$ (i.e., flat) for
the remainder of multiple systems \citep{ps} are
inconsistent with the data unless the true binary fraction
is $F\leq0.60$.  Such low binary fractions are shown to be
improbable. However, mass ratio distributions involving both
a twin component comprising $\sim$40\% of multiple systems
and a power-law component with $\alpha=$ -2 --- -3 (i.e.,
Salpeter-like) are consistent with the data.  }

\item{Secondary star mass distributions comprised of two
components, a power-law component and a ``twin'' component
with $q>0.95\%$, provide good agreement with the data at the
level of $\sim70\%$, marginally better than the best-fitting
single power-law models.  If the fraction of twin systems is
$T_{frac}=25-40\%$ and these secondaries have mean orbital
separations $<\log[r(AU)]>=0.6\pm0.6$, then the data are
consistent with the remainder of the companions being drawn
from the canonical Miller-Scalo -type field star IMF.  This
formalism simultaneously predicts a type Ib/c supernova
fraction of $f_{Ib/c}=$0.30--0.40, consistent with the most
recent estimates.  {\it This two-component description may
allow for the production of massive binary phenomena such as
the large number of double neutron star systems that seem to
be required by the data \citep[see][for a review]{kal07}
while simultaneously permitting the production of low-mass
X-ray binaries at the expected rates.}  }

\item{ The best-fitting binary parameters also predict that
the fraction of core-collapse supernovae that are type Ib/c
supernovae is least 30\% (assuming single stars do not
produce type Ib/c supernovae).  If the actual Ib/c supernova
fraction were 15\%, the allowed binary parameter space would
require binary systems where the two components were all
nearly the same mass (high $\alpha$) and with orbital
separations well above \"Opik's law ($\beta \gtrsim 1$).
Such values, although possible, seem unlikely. More likely
is that the binary observations indirectly support the most
recent supernova observations suggesting that the Ib/c
supernova fraction may be as high as 30--40\%.}

\end{enumerate}

Observational work that better defines the
distribution of binary separations, mass ratios, and  binary
fractions will help to reduce degeneracies in the types of
model constraints used here.  Thereby, such data will allow a more
precise specification of initial binary parameters.  Optical
interferometric observations of companions to massive stars
in the radial range $<20$ $AU$ would help verify that the
\"Opik's Law distribution of separations extends to smaller
orbital distances. Such data may also begin to directly
identify ``twin'' systems with typical separations
$\sim$1--few~$AU$. The continuing Cyg OB2 radial velocity
survey \citep{kiminki} will also measure orbital parameters,
modulo inclinations, for a sizable sample of massive
binaries with a common origin, permitting tighter
constraints on the binary separations and companion masses.

\acknowledgments

We thank the time allocation committees of the Lick, Keck,
WIYN, and WIRO Observatories for generous observing
allocations and Stan Woosley for making this project
possible.  We acknowledge helpful conversations with Peter
Conti, Steve Vogt, Laura Penny, Bob Mathieu, Doug Gies, and
Ed van den Heuvel.  We acknowledge Dan Kiminki for
his assistance with the Cygnus OB2 survey 
data in the preparation of this manuscript.
We thank the referee, Ronald Webbink,
for his substantial investment of time on comments that greatly
improved this paper.  We are grateful for support from the
National Science Foundation through the Research Experiences
for Undergraduates (REU) program grant AST-0353760 and
through grant AST-0307778.  The work of C.~F. was in part
under the auspices of the U.S.\ Dept.\ of Energy, and
supported by its contract W-7405-ENG-36 to Los Alamos
National Laboratory and by National Science Foundation under
Grant No. PHY99-07949.

{\it Facilities:} \facility{WIRO ()}, \facility{WIYN ()}, \facility{Shane ()}, 
\facility{Keck:I ()}

{}

\clearpage

\clearpage

\clearpage

\clearpage

\clearpage

\clearpage

\clearpage

\clearpage

\clearpage

\clearpage

\clearpage

\clearpage

\clearpage

\clearpage

\clearpage

\clearpage

\end{document}